\RequirePackage{snapshot}
\documentclass[final,3p,times]{elsarticle} \biboptions{sort&compress} \pdfoutput=1
\usepackage{graphicx}
\graphicspath{{./figs/}, {./figs/BCube/}, {./figs/FatTree/}, {./figs/MDCube2D/}, {./figs/Biography/}}
\DeclareGraphicsExtensions{.pdf,.jpg,.jpeg,.png}

\usepackage[cmex10]{amsmath}
\usepackage{mathtools}
\usepackage{amssymb}
\usepackage{algorithmicx}
\usepackage[ruled]{algorithm}
\usepackage[noend]{algpseudocode}

\usepackage{array}

\usepackage{amsmath}
\usepackage{amssymb}
\usepackage{graphicx}
\usepackage{array}
\usepackage{multirow} 
\usepackage{rotating}

\usepackage[labeled,resetlabels]{multibib}
\newcites{S,P}{Supplementary References,	Postscript References}
\usepackage{float}

\usepackage[usenames,dvipsnames]{xcolor}
\usepackage{float} 
\usepackage{ulem}
\usepackage{color}
\usepackage{soul}
\sethlcolor{GreenYellow}
\renewcommand{\hl}[1]{#1}

\usepackage{colortbl}

\usepackage{afterpage} 
\usepackage{setspace}

\usepackage{picins} 
\usepackage{bm}

\makeatletter
\let\ftype@table\ftype@figure
\let\ftype@algorithm\ftype@figure
\makeatother

\newcommand{\LineComment}[1]{\State \(\triangleright\) {#1}\hfill}
\algblock{ParFor}{EndParFor}
\algnewcommand\algorithmicparfor{\textbf{parfor}}
\algnewcommand\algorithmicpardo{\textbf{do}}
\algnewcommand\algorithmicendparfor{\textbf{end\ parfor}}
\algrenewtext{ParFor}[1]{\algorithmicparfor\ #1\ \algorithmicpardo}
\algrenewtext{EndParFor}{\algorithmicendparfor}

\usepackage{spreadtab}
\STautoround*{2} 
\usepackage{numprint}
\npthousandsep{,~}
\npthousandthpartsep{}
\npdecimalsign{.}

\makeatletter
\renewcommand{\maketag@@@}[1]{\hbox{\m@th\normalsize\normalfont#1}}\makeatother

\def\por1{\partial}

\newcolumntype{S}{>{\centering\arraybackslash} m{.4\linewidth} }
\newlength{\figuresintablefour}
 \newlength{\figuresintablesix}
 \newlength{\figuresintableeight}
 \newlength{\figuresintablenine}
\makeatletter
  \newcommand\tinyv{\@setfontsize\tinyv{5pt}{7}}
\makeatother

\newlength{\hspacephantom}
\usepackage{url}
\usepackage{cooltooltips}

\usepackage[strings]{underscore}

\journal{Journal of Network and Computer Applications}
\begin{document}

\begin{frontmatter}

\title{\large A Monte-Carlo Approach to Lifespan Failure Performance Analysis of the Network Fabric in Modular Data Centers}

\author{\small Reza~Farrahi~Moghaddam,$^{1,2,*}$ Vahid Asghari,$^{1}$ Fereydoun~Farrahi~Moghaddam,$^{1,2}$ Yves~Lemieux,$^{2}$ \textnormal{and} Mohamed~Cheriet$^{1}$}
\address{\small $^{1}$ Synchromedia Lab, ETS (University of Quebec), Montreal, QC, Canada H3C 1K3}
\address{\small $^{2}$ Ericsson Research - Cloud Technology, Ericsson Canada Inc, Montreal, QC, Canada H4P 2N2}
\address{$^{*}$ Corresponding Author: \url{imriss@ieee.org}, LinkedIn: \url{https://www.linkedin.com/in/rezafm}      }
\address{DOI: 10.1016/j.jnca.2017.03.015, \url{http://dx.doi.org/10.1016/j.jnca.2017.03.015}, Accepted: Mar 14$^{th}$, 2017}

\date{Accepted: Mar 14$^{th}$, 2017}

\setstretch{1.25}
\begin{abstract}
Data centers have been evolved from a passive element of compute infrastructure to become an active, core part of any ICT solution. In particular, modular data centers (MDCs), which are a promising design approach to improve resiliency of data centers, can play a key role in deploying ICT infrastructure in remote and inhospitable environments in order to take advantage of low temperatures and hydro- and wind-electric capabilities. This is because of capability of the modular data centers to survive even in lack of continuous on-site maintenance and support. The most critical part of a data center is its network fabric that could impede the whole system even if all other components are fully functional,
\hl{assuming that other analyses has been already performed to ensure the reliability of the underlying infrastructure and support systems.}
In this work, a complete failure analysis of modular data centers using failure models of various components including servers, switches, and links is performed using a proposed Monte-Carlo approach. The proposed Monte-Carlo approach, which is based on the concept of snapshots, allows us to effectively calculate the performance of a design along its lifespan even up to the terminal stages. To show the capabilities of the proposed approach, various network topologies, such as FatTree, BCube, MDCube, and their modifications are considered. The performance and also the lifespan of each topology design in presence of failures of their components are studied against the topology parameters. 
\end{abstract}


\end{frontmatter}

\setstretch{1.5}

\section{Introduction}
\label{sec_introduction}
Traditionally, the data center (DC) term has been used as a reference to the warehouse-scale (up to tens of thousands of high-performance servers as well as the associated networking, storage, cooling, and support equipment) facilities that centralize a company's ICT-related operations and equipment. Nowadays, in a networked society \cite{Ericsson2014b}, DCs have rather become a vital part of seamless operation of many daily activities. Consequentially, the reliability and availability (uptime) of DCs and their services have been a top priority for many companies and organizations. In particular, Virtualization, Everything as a service (XaaS),\footnote{For example, Infrastructure-as-a-Service (IaaS).} and open source paradigms have enabled the DCs and, more generally, compute/network/content (CNC) delivery solutions to operate at the level of {\em utilities} (with the new responsibilities \cite{FCCNetNeutrality2015a}). 

Low levels of reliability and dependability, such as those related to unreachability, traffic choking at link as well as at component (switches) levels, and also misconfiguration could highly affect the Return on Investment (RoI) of a DC. In particular, having a reliable DC network fabric with a scalable architecture design is a critical concern. 
\hl{However, it is worth mentioning that a failure of a data center is not limited only to its ICT-level components including the network fabric. In particular, a failure in the infrastructure and support system of a data center, including the power supply system and the cooling system, could result in a failure of the whole operation. From here on we assume that another analysis has been performed to ensure the reliability of the support systems, and we only focus on the ICT-level components namely servers, switches and links. It is also worth mentioning that because of aggregated nature of the output of the support systems, straightforward solutions such as replication of such systems could be used to increase their resiliency with respect to failures.}
By definition, in a DC network, reliability is the ability to perform and deliver the required functions and services for a specified period of time and under some conditions declared in service level agreements (SLAs). Scalability in terms of expanding hardware and also installing DCs at unsecured or unmonitored places could be identified as two important challenges for reliability. To address these challenges, DCs with a scalable architecture are designed using purpose-engineered modules that at the same time reduces costs of operation and administration. This modular data center (MDC) approach is also becoming more popular for large-scale DC operations \cite{Ramfitt2011}. MDCs can be imagined as set of independent, self-directed compute modules (such as cabinets or physical containers) that are built in standardized sizes, such as that of overseas shipping containers, hosting a large number of servers \cite{Hamilton2006}. 

It is worth noting that software level management of a DC is a big challenge by itself. However, designing an agile, robust, and scalable DC hardware (at both topology and component levels\footnote{The disaggregated hardware approach provides flexibility at the level of components (nodes) \cite{Ericsson2015e,Mainstay2016}. The flexibility and resilience at the topology level is the focus of this work.}) is the fundamental requirement that affect all high-level management layers. This is more critical for ship-and-deploy (S\&D) MDCs that seem to be promising solutions for fast deployment of resilience compute resources at any place and any time. These self-contained MDCs could be deployed in remote areas where only the Electricity, Water, and Broadband connectivity (EWBc) infrastructure is in-place, but there will be no human support/maintenance/intervention available in short-term periods of time. This is highly interesting, and is especially along with strategic north plans in Canada at both provincial and national levels \cite{Berger2010,Fournier2012,Fiser2012}, which target Canada's natural resources, especially untouched hydro-electric and wind-electric sites in the North \cite{Hofman2009}, among other possibilities.

There are a number of work addressing the reliability concern in the DC network, and a few DC network architectures have been investigated in order to provide a better reliability and scalability in the context of MDCs (Please see the Related Work section, Section \ref{sec_Related_Work}, for more details). 
It is worth noting that the reliability analyses provided in these studies, such as in \cite{Couto2012}, have been based on removing a few components according to the failure relations in order to evaluate the effect of each component type's failure on the reliability of the DC network topology. In other words, although these studies investigate the {\em relative} importance of various components, they could not provide the whole picture of degradation and decrease in reliability of the DC along time. In contrast, the operation of MDCs highly depends on such a time-dependent analysis.

\begin{figure*}[tbh!]
	\begin{tabular}{cc}
		\fbox{\includegraphics[height=7cm]{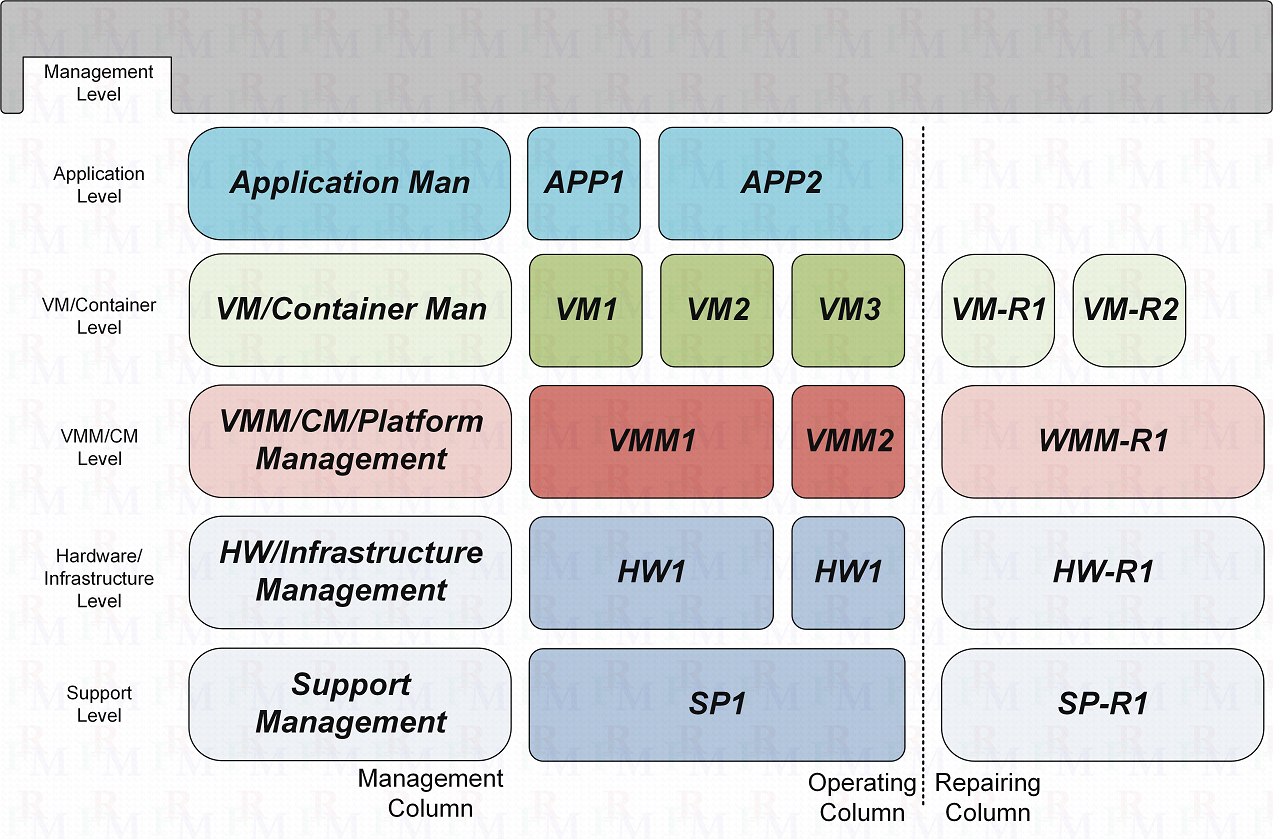}} & \fbox{\includegraphics[height=7cm]{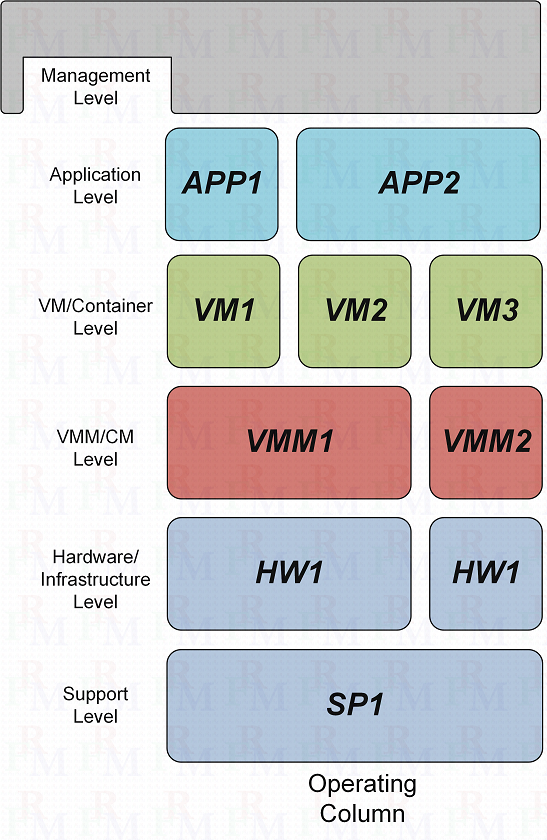}} \\
		(a) & (b)
	\end{tabular}
	\caption{a) A schematic stack of components layers. 		b) The simple case of MDCs in this work, which focuses on the failure of the compute components. The maintenance, support and management components and layers are excluded in this case.}
	\label{f-system-schematic1}
\end{figure*}

Second, although we can conclude a sigmoid-like reliability behavior for a system similar to those of its physical components, the system availability cannot be bounded with only this information. In practice, a system would go through several unavailability episodes (time intervals) during its lifetime because of repairable failures that happen at higher levels/layers of the system (above the physical level/layer). A simplified schematic of a typical system is shown in Figure \ref{f-system-schematic1}(a). `Software Aging' phenomena in the software levels, such as Virtual Machine Managers (VMM) and Container Managers (CM) level, would eventually lead the system, and also the individual applications hosted on it, to fail or degrade to an unacceptable level of Service Level Objectives (SLOs). These episodes could be treated with manual or automatic interventions that can be easily performed remotely. Analysis of the (sub)-system availability and the impact of individual components, such as VMMs, is beyond the scope of this work. However, the lifetime analysis provided herein could be of great interest in planning high availability designs, since the availability of a system cannot go beyond its reliability lifetime. 

In this paper, we propose a new analysis approach based on individual component failure profile described in terms of cumulative distribution functions (CDFs). The proposed analysis approach, which is based on Monte-Carlo methodology, calculates  the reliability picture of the system under analysis along with several performance indicators. These indicators are proposed to provide insights in terms of profitability of a solution from various perspectives. The details are provided in the subsequent sections. In fact, up to our knowledge, there exists no study on comparing DC network topologies in terms of failures on DC components, such as servers, switches, physical links and so forth, while taking into account the operational life-time of the DC. To be precise, we use the proposed Monte-Carlo approach to conduct a series of experiments in which a fault-free topology is initiated at time $t=0$, and then its behavior is evaluated along the time. The performance profile of the topology is then generated along time with respect to one or a few parameters of the topology.

The paper is organized as follows. Section \ref{sec_notations} provides a short list of notations used in the rest of the paper. The proposed Monte-Carlo methodology to analyze the failure performance of data centers along their lifespan is introduced in section \ref{sec_MC_Analysis} \hl{along with the experimental setup described in Section} \ref{sec_Experimental_Setup}. Next, the performance metrics are presented in section \ref{sec_Perf_Metrics}. Then, the performance of the three major topologies and their generalizations are studied in section \ref{sec_Res_Discussions}. This is followed by the related work and some conclusions presented in sections \ref{sec_Related_Work} and \ref{sec_conclusion}, respectively. In \ref{sec_Tanh}, the proposed Tanh-Log cumulative distribution function is introduced and fitted using the data from a real system database. Then, in \ref{sec_MDCs} and \ref{sec_star-topo}, the three main topologies and their modifications that are considered in this study are presented and discussed.

\section{Notations and Definitions}
\label{sec_notations}
In this section, some of the graph theory notations used from here on are presented. 
\begin{description}
\item[Directed Graph $G^{d}(V,E,w)$:] A directed graph with vertex set $V$ of $n=||V||$ nodes along with the set $E$ of its $m=||E||$ active edges and their associated weights $w$. A vertex is denoted either by $v_i$ or by its index $i$ (which is equivalent to its resource ID): $v_i \in V$. We may also causally use $i \in V$ as a short notation. A typical directed edge is shown by $e^{d}_{i,j}$, where $i$ and $j$ are the the start and end vertexes. Also, $w^{d}_{i\rightarrow j}\left(=w^{d}\left(e_{i,j}\right)>0\right)$ denotes the associated positive weight of the edge $e^{d}_{i,j}$. We assume $w>0$ considering the fact that in this work the weights actually represent the bandwidth of a link in a data center network. 
\item[Undirected Graph (Network) $G(V,E,w)$:] An undirected graph with vertex set $V$ along with the set $E$. For each edge $e_{i,j}$ there are equivalently two edges: $e^{d}_{i,j}$ and $e^{d}_{j,i}$, which means a graph $G$ could be equivalently built using a $G^{d}$ that has two edges between each pairs of connected nodes. Considering the fact that in the data center networks and fabrics the links are usually symmetric, we use a single parameter $w_{i,j}$ to represent both equal bandwidth of a link between nodes $i$ and $j$. 
\item[Node $v_i$:] A node $v_i$ of a data center network could be a compute node (a server) or a network node (a switch). The compute nodes focus on the compute intensive functions, while the network nodes perform better for the non-local data processing (including transport, routing, and switching, among others). Considering the introduction of white-box hardware and also open hardware, the differentiating factors between compute and network nodes is counting down to criteria such as design preference, performance, and cost. For example, in a BCube topology, the compute nodes also participate in the networking functions (please see Section \ref{sec_star-topo}).
\item[State of a node $\mathcal{s}_i(t)$:] The state of a node with respect of failure at a specific time $t$. In this work, the state values are binary from a set of $\left\{\text{Operational}, \text{Failed}\right\}$.
\item[State of a link $\mathcal{s}_{i,j}(t)$] The state of the link between two nodes $i$ and $j$ with respect of failure at a specific time $t$. The set of possible values is $\left\{\text{Operational}, \text{Failed}\right\}$.
\item[State of the graph $\mathcal{S}_t$] The state $\mathcal{S}_t$ of a graph $G$ is an instance of $G(V,E,w)$ at a specific time $t$, in which all vertices and links have been attributed with their specific state value.
\item[Trajectory $\mathcal{T}$] A Trajectory $\mathcal{T}$ is a sequence of states of ``a'' particular graph $G$ from the initial time $t=t_0=0$ till the current time $t=t_1$: $\left(\mathcal{S}_t\right)_{t=0}^{t=t_1}$. For the purpose of simplicity, we omitted the graph in the notation of the trajectory. It is worth mentioning that if a constant graph $G$ is repeatedly exercised from the same initial state at time $t=0$, it may follow different trajectories in each one of those exercises. This means that the graph is not allowed to switch between trajectories (although at some points along time, the trajectories may become nondifferentiable). Considering the fact that we are interested in improving the overall performance of a large number of similar data centers (an ensemble of data centers), the performance metrics (to be mentioned in Section \ref{sec_Perf_Metrics}) are statically evaluated on the ensemble of the trajectories at each specific time, and then are augmented to generate the performance profile.
\item[Snapshot $S$] A snapshot is a particular state $\mathcal{S}$ of a graph $G$ at a time $t$: $S=\left(\mathcal{S}, t\right)$. It is a tuple of a state $\mathcal{S}$ and a time instance $t$. Implicitly, we assume that the graph has been exercised from $t=0$ with the initial state. It could be argued that a snapshot $S$ is the state at the current time $t$ of its associated trajectory with the assumption that the trajectory is not {\em retrievable}. In terms of the performance of an ensemble of data centers, the statistics could be evaluated on a large-enough set of snapshots at a specific time. In the case of Modular Data Centers (MDC) of this paper (to be discussed in Section \ref{sec_MDCs}), which do not undergo maintenance and repair operations during long time intervals, the set of snapshots at a specific time could be directly populated based on the initial state, the partial differential function (PDF) of individual components of the data center network (nodes and links), and the target time, without requiring to generate the associated trajectories.
\end{description}

\begin{figure}[tbh!]
\centering
\begin{tabular}{c}
\includegraphics[width=10cm]{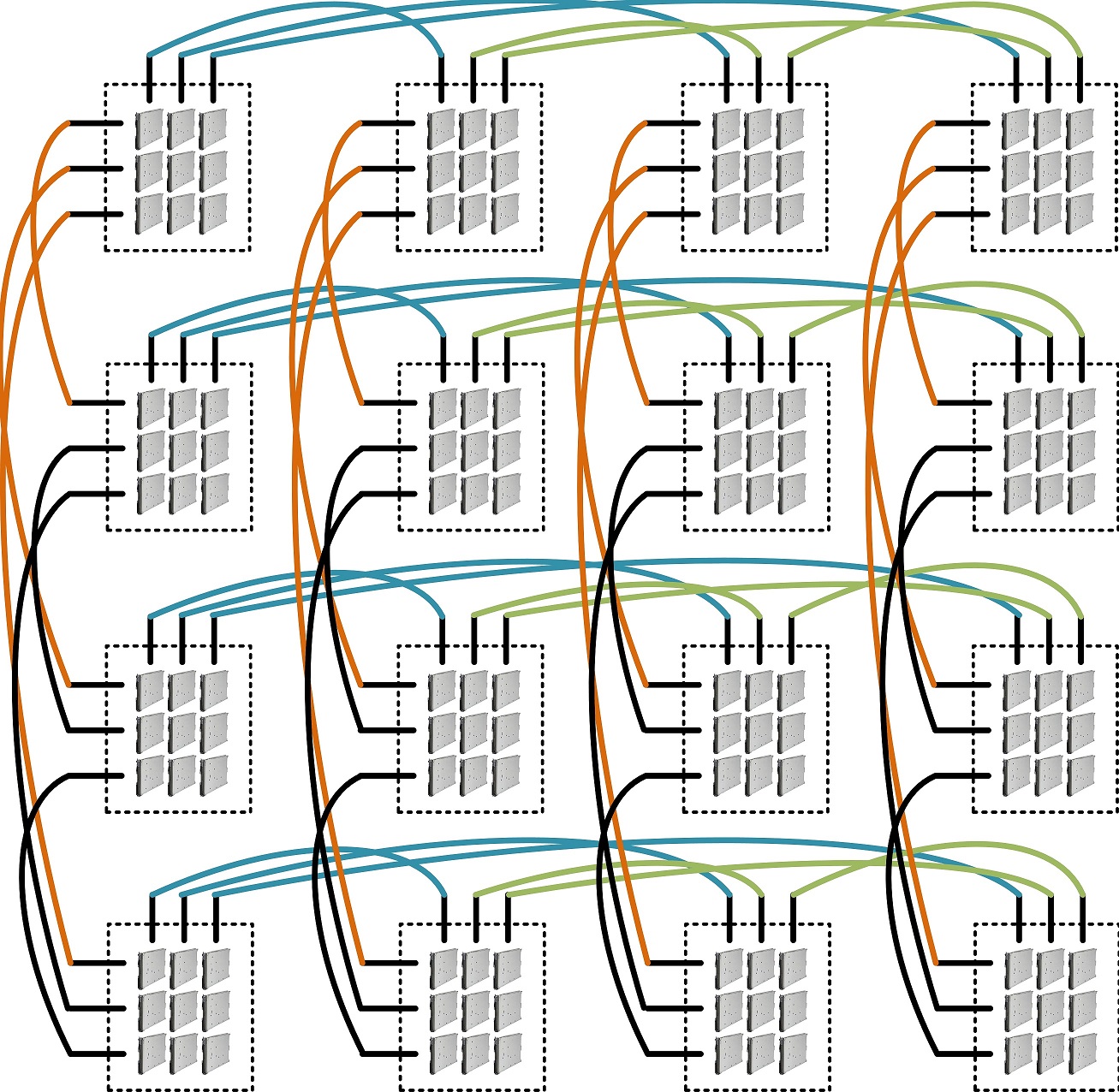}
\end{tabular}
\caption{An illustrative example of the MDCube2D topology. There are four containers per row, each one a BCube (level-1) with 2-port servers and switches arranged in three BCube (level-0) cubes.}
\label{f-mdcube-schematic1}
\end{figure}

\section{The Proposed Snapshot-based Monte-Carlo (MC) Analysis of Data Center Topologies}
\label{sec_MC_Analysis}
\begin{figure}[t!]
	\centering
	\begin{tabular}{c}
		\includegraphics[width=12.5cm]{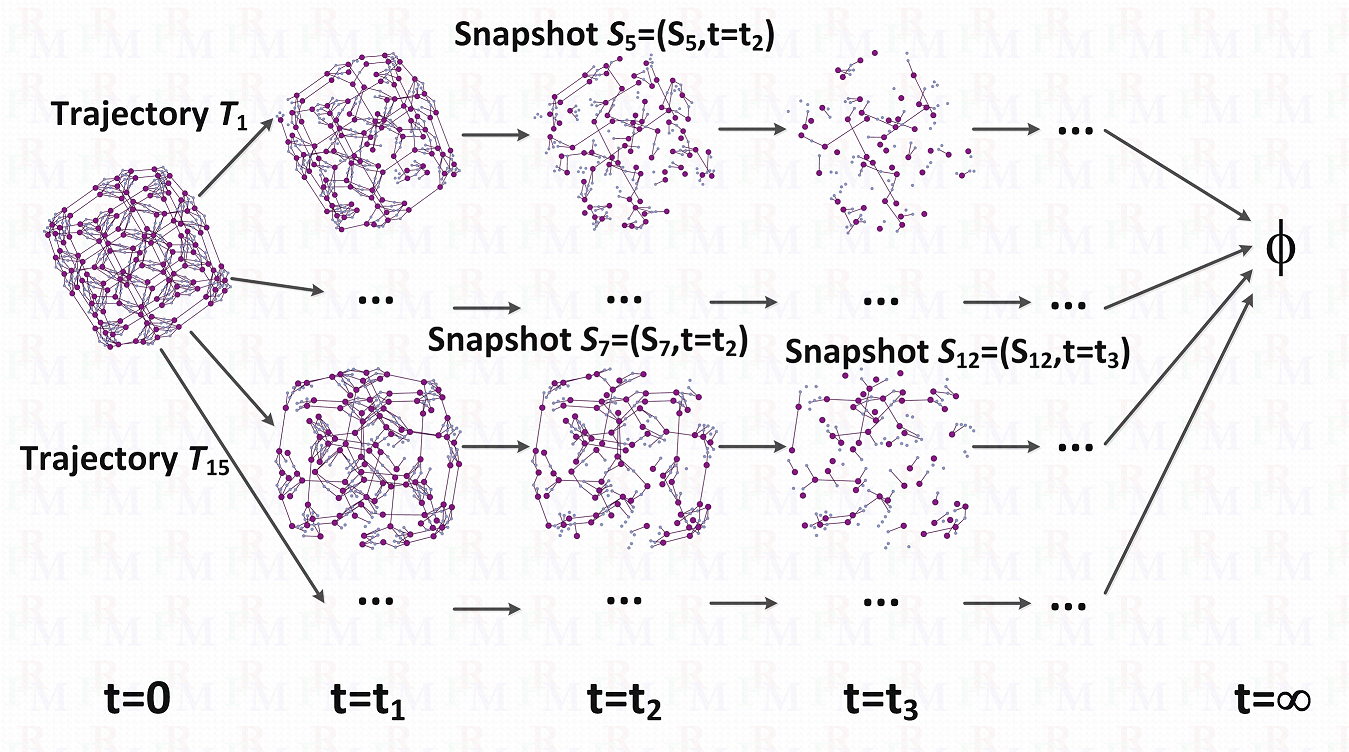}
	\end{tabular}
	\caption{A typical and illustrative example of evolution of an MDC topology along the time in presence of failure. The MDC topology is an MDCube2D$^{E}$ with 256 servers and 128 switches, which are shown in blue and red nodes, respectively. The system can fall in one of many possible trajectories two of them are shown in the figure. From each trajectory only one snapshot would be eligible to be considered. At $t=\infty$, all trajectories reach to a common state in which all nodes and links have failed (denoted by the empty set $\phi$ in the figure).}
	\label{f-snapshot1}
\end{figure}

Let's consider an MDC with specified topology and components. It is assumed that the system is fully operational at $t=0$. Therefore, there is only one state of the system is possible at time $t=0$. For any $t>0$, the set of reachable system's states is equal to the set of all possible system's states. However, the probability of having any of those states could be different depending on $t$ and the degree of failure in the system. In the proposed MC approach, a set of snapshots for each $t$ is taken from its set of possible system's states according to their probabilities. These sets of snapshots along the time provide a statistical picture of the performance of the specific topology of the MDC measured using the performance metrics aggregated on the snapshots. For each component, the CDF of failure is used along our in-house random number generator in order to determine its state in terms of being operational or failed. This process is repeated for all the components, and at the end a snapshot of the system at that specified instance of time is created. We have used a size of 100 snapshots per time instance in this study. A schematic diagram illustrating the concepts of paths and snapshots is provide in Figure \ref{f-snapshot1}.\footnote{A sample video of the evolution of a MDCube2D$^{E}$ topology along one of its trajectories can be viewed at this link: 
\cooltooltip{Trajectory}{A link to a trajectory}{http://youtu.be/Y0G7xEiY4fk}{Watch a trajectory on the website}{http://youtu.be/Y0G7xEiY4fk}.
} Each graph represents a snapshot of a MDC at its associated time. Only two of many possible trajectories are drawn. At $t=0$, all trajectories have the same snapshot, from which they diverge to their own path. The MDC used is a MDCube2D$^{E}$ topology\footnote{For the detailed description of the topology, please see \ref{sec_star-topo}.} with 256 servers and 128 switches. The servers are represented by blue nodes, while red nodes correspond to the switches. There is a convergence at the end of life along all trajectories and paths. It is worth noting that in the proposed analysis approach, as mentioned before, the snapshots of a time instance $t$ are not built based on the snapshots taken from its preceding instance $t^{-}$.\footnote{$t^{-}$ is any indistinguishable time instance {\em before} $t$.
} In other words, each snapshot should have its own associated trajectory.

\hl{The snapshots are assumed to be trajectory-independent, i.e., a snapshot is created by populating the state of every component (server, switch, and link) using their distribution functions, random generators, and the time. When we move from one time-instance to the next one, the new snapshots are generated independently from the previous time instance. In a complex system with many components and over long periods of time, th number of possible system-level states would be enormous at any time instance, and therefore the number of possible trajectories would be high. The nature of modular data centers, in which a component state only depends on its distribution function and the period of time passed, allows us to perform the snapshot-based analysis and ignore the trajectories. Hypothetically, we can assume that every snapshot belongs to at least one trajectory. This fact has been shown in Figure 3. However, it should not be interpreted that in the proposed analysis the snapshots are calculated using trajectory.}

\begin{figure}[t!]
	\centering
	\begin{tabular}{c}
		\includegraphics[width=12.5cm]{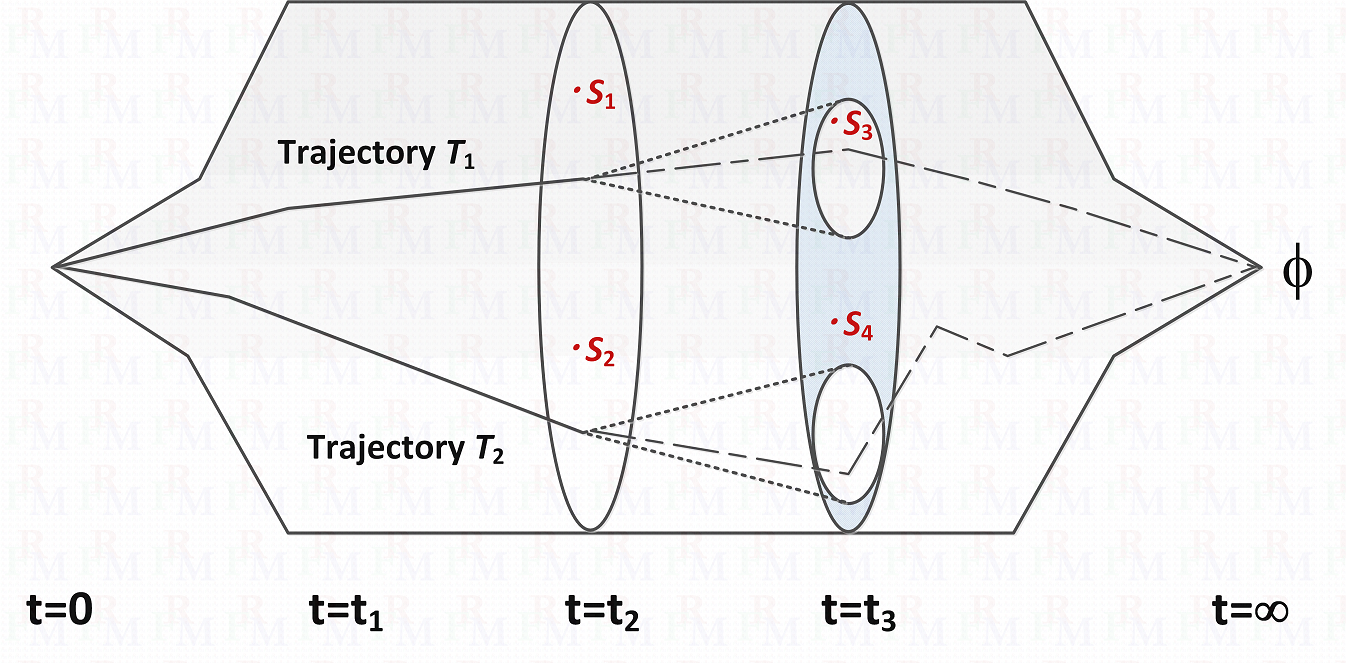}
	\end{tabular}
	\caption{\hl{A illustration of evolution of state space along the life time of an MDC topology. At each time instance, the sub-space of the possible states is shown by a circle. The proposed snapshot-based approach would outperform the trajectory-based approaches in terms of uniformity of sampling at each time instance becuase of its independence from the performance at the preceding time instances.}}
	\label{f-snapshot2}
\end{figure}

\hl{Although it may be argued that considering the limited number of samples considered in each analysis, a trajectory-based analysis would perform better or the same as the proposed snapshot-based analysis, we argue that considering the large number of possibilities, the sampling is always a sub-sampling, and the proposed snapshot-based analysis would give a better-distributed and ``objective'' sampling because, in contrast, any trajectory-based analysis would suffer from its high dependency on the actual numbers that are used for the few first time instances (these few first samplings would limit the trajectory ``cone'' regardless of the quality of the random numbers that are used for the rest of the time instances). In other words, engineering efficient trajectories that traverse the long life time of a modular data center would be a great mathematical challenge by itself, while the snapshots' performance is mainly limited by the number of snapshots used. Figure} \ref{f-snapshot2} \hl{schematically illustrates this fact. At the time instance $t=t_2$, the sub-space of the possible states is shown in the form of a hyper-circle, and also only two trajectories are shown for the purpose of clarity of the graph. It is straightforward to see that the possible states at $t=t_3$ that can be reached from the two trajectories' landing state in the sub-space at $t=t_2$ are highly limited to the ``cones'' that start from those landing points (illustrated by two small circles in the sub-space of $t=t_3$). The inaccessible part of the sub-space in the $t=t_3$ is marked by crossed hashed pattern. In contrast, the proposed snapshot-based approach is expected to perform a uniform sampling in each of the sub-spaces, and therefore its outcome is not affected by the previous samplings performed in the preceding time intervals. For example, in Figure} \ref{f-snapshot2}, \hl{the snapshot $S_4$ is placed in the crossed-hashed part at $t=t_3$ which would be impossible for trajectory-based approaches to reach it.}

The Algorithm \ref{alg_Monte_Carlo_1} provides the summary of the proposed Monte-Carlo approach. The algorithm receives an input which includes the topology family and its parameter values in addition the time interval and the distributions. Then, it continues with generating snapshots at a series of time instances selected along the time interval. After generating the snapshots at a specific time instance, the statistics and then the associated performance metrics are calculated. The performance metrics along the time series is collected in order to perform statistical post-analyses, such as that of confident interval analysis. It is worth mentioning that the proposed Monte-Carlo algorithm is of a linear complexity with respect to the time interval thanks to using snapshots because of no-repair specifications of the MDCs.

\begin{algorithm}[!htbp]
	\small
	\caption{A Snapshot-based Monte-Carlo algorithm for MDCs.}
	\label{alg_Monte_Carlo_1}
	\begin{algorithmic}[1]
		\Procedure{$\text{PerformanceProfile} = \mathbf{SnapshotMC}$}{TopologyName, TopologyParameters, $t_\text{final}$, DeviceDistributions, PerformanceMetrics}
		\LineComment{\emph{Initialize the network $G$ at $t=0$ (All nodes and links are set to `up' state).}}
		\State $G \gets$ $G_0$;
		\LineComment{\emph{The algorithm marches the network $G$ along the time from $t=0$ till $t=t_\text{final}$.}}
		\State $t  \gets 0$;
		\State $S_0 \gets (G_0, t)$;
		\State $\text{Performance}_0 \gets \text{PerformanceMetrics}(S_0)$;
		\Repeat
			\LineComment{\emph{For each new time instance:}}
			\State $t \gets t + \delta t$; 
			\LineComment{\emph{Each snapshot at $t$ is preferably analyzed independent from the others.}}			
			\ParFor {$s = 0, \cdots, n_\text{snapshot}$}
				\LineComment{\emph{Generate the snapshot $S_{t,s}$.}}			
				\State $S_{t,s} \gets$ Using the DeviceDistributions and $G_0$;				
				\LineComment{\emph{Calculate the statistics and also the performance metrics of the snapshot.}}
				\State $\text{Performance}_{t,s} \gets \text{PerformanceMetrics}(S_{t,s})$;
				\LineComment{\emph{Store the time, the performance metrics and statistics of the snapshot.}}
			\EndParFor
			\LineComment{\emph{Calculate the statistics of the performance metrics at $t$ across all snapshot data.}}	
			\State $\text{PerformanceProfile} \gets \text{PerformanceProfile} \oplus \text{Performance}_{t}$;		
			\Until{$t \geq t_\text{final}$.}
		\LineComment{\emph{Return the calculated performance metrics and statistics.}}
		\State {\bf Return} $\text{PerformanceProfile}$
		\EndProcedure
	\end{algorithmic}
\end{algorithm}

\section{Experimental Setup}
\label{sec_Experimental_Setup}
In this section, the experimental setup including the performance metrics are presented and listed. The results of the experiments are then provided in the next section, Section \ref{sec_Res_Discussions}, along with the discussions.

\subsection{Setup}
\label{sec_Setup}
To show the capability of the proposed Monte-Carlo approach, a series of experiments are considered. In each experiment, a fault-free topology is initiated at time $t=0$, and then its behavior is evaluated along the time. In order to have a statistically meaningful result at each time instance, the state of the evolving topology is populated many time ($n_\text{snapshot}=100$, in this setup) by generating snapshots of the topology at that time instance with respect to the initial topology at $t=0$. Having the ($n_\text{snapshot}$ snapshots generated for a time instance, the associated statistical data is calculated and then is used to calculate the performance metrics of that time instance. The performance profile of the topology is then generated concatenating the calculated matrices along time. Considering the characteristics of the topology, the performance metrics and profiles are analyzed with respect to one or a few parameters of the topology. In the same way, the families of topologies (such as FatTree family or BCube family) are compared against each other while we try to keep the comparison fair (by enforcing the same number of server nodes for each one).

At the component level, we have three types of devices: Servers, switches, and basic links. The failure profile of the servers has been observed to resemble the the well-known Weibull distribution \cite{Kondo2010}. we have fit this distribution on the empirical {\em lanl05} database \cite{Schroeder2010} to have the actual failure distribution model for the servers. Also, we have used a proposed {\em Tanh-Log} distribution for the purpose of fitting on the same data. The details of the failure models and the fitted coefficient are provided in \ref{sec_Tanh}. The proposed {\em Tanh-Log} distribution provided a better goodness of fit (GOF). Therefore, we use the latter distribution in our experiments. It is worth mentioning that regardless of what distribution is used in the failure models, the proposed Monte-Carlo approach provide us with the performance metrics in the validity range of the distribution used. Therefore, for the other two types of devices, i.e., the switches and links, we used the same dataset and scale the time values with a factor of 5 and 10, respectively, before applying the fit function. Although this approximation may not regenerate an accurate picture of a real data center, it allows us to show the capability of the proposed Monte-Carlo approach in calculating the performance metrics. In the future, we will study the differentiating impact of the Tanh-Log distribution versus the Weibull distribution on various dataset in order to highlight potentials of the proposed distribution. 

\hl{Although in this work we assume that all components behave independently from each other, it is worth mentioning that some components such as switches may correlatively fail in highly tight time intervals because of software faults of their identical firmware} \cite{Crowell2004,Stern2016}. \hl{We will address such cases in a future work in which highly correlated failures at the component level are modeled by introducing subgroups in the main three groups of components (servers, switches, and links). For each subgroup a different time-dependent distribution function of failure is assigned.}

At the level of topology selection, we consider three well-known topologies in our setup: FatTree \cite{Fuentes2012,Navaridas2010}, BCube \cite{Guo2009}, and MDCube \cite{Wu2009}\footnote{We particularly consider the two-dimensional version, i.e., MDCube2D.}. In addition, in a specific use case of 3072 servers, a series of modified version of these topologies are considered. The modified version, denoted as FatTree$^{E}$, BCube$^{E}$, and MDCube2D$^{E}$, are introduced in \ref{sec_star-topo}. These modified topologies are intended to allow utilization of high-port-number switch while relaxing the relation between the number of port of the actual switches and the number of servers placed in each unit of the topology architecture. The details of these modified topologies could be found in the appendix. Again, the proposed Monte-Carlo approach calculates the performance metrics regardless of the actual topology selected in each experiment.

\subsection{Performance Metrics}
\label{sec_Perf_Metrics}
All metrics in this section are defined on a single system snapshot, i.e., an instance of the system with some of components failed. The process to generate the snapshots and also the procedure to aggregate the metrics calculated on the set of snapshots have been discussed in Section \ref{sec_MC_Analysis}.

Assuming that there is no software-related constraint, the maximum pool of resources of a modular data center is the maximum number of connected servers, which we call it the largest pool of the data center. At the time $t=0$, the largest pool includes all the servers. We consider two performance metrics with respect to pool size: These metrics, defined for each snapshot, are the {\em maximum} and {\em average} relative sizes of the connected components after failures in DC devices, which are defined in analogy to the metrics used in the studies on reliability in complex networks \cite{Albert2000a,Fortunato2010}. The relative size of the largest connected component to the total number of servers in all pools, called Maximum relative size, is denoted $\text{RS}_{\max}$: 
\begin{align}
	\small
	RS_{\max}\left( t_1 \right) &= \mathop {\max}\limits_{i \in I} \left\{ RS\left( t_1, i \right) \right\} = \frac{\mathop {\max}\limits_{i \in I} \left\{ n_{{\rm{srv}},i}\left( t_1 \right) \right\} }{\sum_{i\in I} {n_{{\rm{srv}},i}\left( t_1 \right)}}  
	\label{Metrics_1}\
\end{align}
where $n_{{\rm{srv}},i}(t_1)$ denotes the number of servers available in the $i^{th}$ connected components at time $t_1$. It is worth mentioning that although various snapshots are possible for any time instance $t_1$, we use the same time value to index the snapshots in the metrics for the purpose of simplicity of notation. 

\subsubsection{Maximum absolute relative size ($\text{ARS}_{\max}$)}
Absolute relative size (ARS) of a connected component is defined as the ratio of the number of servers in that connected component to the total number of servers of initial state. 
The Maximum absolute relative size, $\text{ARS}_{\max}$, is defined as the maximum of all individual ARS values of all system's connected components at time $t_1$:
\begin{align}
	\small
	\text{ARS}_{\max}\left( t_1 \right) &= \mathop {\max}\limits_{i \in I} \left\{ ARS\left( t_1, i \right) \right\} = \frac{\mathop {\max}\limits_{i \in I} \left\{ n_{{\rm{srv}},i}\left( t_1 \right) \right\} }{N_{\rm{srv}}}  
	\label{Metrics_2}\
\end{align}
where $N_{\rm{srv}}$ denotes the total number of servers in all the connected components at time $t_0$, i.e., $N_{\rm{srv}}=\sum_{i\in I}{n_{{\rm{srv}},i}\left( t_0 \right)}$, where $t_0$ denotes the starting time of the MDC utilization. In this work, $t_0=0$.

\subsubsection{Total absolute relative size ($\text{ARS}_{\text{total}}$)}
For a single leaser (tenant) of the MDC, the maximum absolute relative size ($\text{ARS}_{\max}$) provides the maximum service level that can be provided to that leaser (for example, an enterprise). However, DCs are most likely operated in {\em multi-enterprise}\footnote{	Also, it is informally spelled as multi-tenant.} configurations, and therefore, the total profitability and revenue of the DC cannot be estimated from $\text{ARS}_{\max}$. To address this need, we introduce the Total absolute relative size metric, $\text{ARS}_{\text{total}}$, that counts the total number of surviving servers in the {\em big-enough} pools to the initial number of servers. To define the profitable pools, a minimum threshold for the the number of servers of a pool is considered, and all pools with less that this threshold size are ignored in the calculations of $\text{ARS}_{\text{total}}$. The minimum threshold is denoted $n_{\rm{srv,req}}$:
\begin{align}
	\small
	\text{ARS}_{\rm{total}}\left( t_1, n_{{\rm{srv,req}}} \right) &= \frac{\mathop {\sum}\limits_{i \in I} \left[ \frac{ n_{{\rm{srv}},i} \left( t_1 \right)}{n_{{\rm{srv,req}}}} \right] \times n_{{\rm{srv,req}}} }{N_{\rm{srv}}}.  \label{Metrics_3}\
\end{align}

\subsubsection{Maximum network flow (${\text{Flow}}_\text{max}$)}
Considering increasing demand to process bigger volumes of data and also physical limits of maximum data transfer rate on copper or optical links, multi-path data transfer has attracted a considerable attention, and with ongoing efforts on SDNs and other initiatives to ease the restrictions of software components, the multi-path data transfer in DCs would reach the hardware upper limit of multi-path transferring in near future. Here, we introduce a metric to evaluate the quality of a topology in terms of multi-path data transfer considering failure in its components. 
The maximum network flow, ${\text{Flow}}_{\max}$, of the $i^{th}$ connected components at the time $t_1$ is defined as follows:
\begin{align}
	\small
	\text{Flow}_{\max} \left( t_1, i \right) = {\rm{mean}} \left( \left\{ \text{Flow} \left( {{\rm{srv}}_l,{\rm{srv}}_m} \right)  \Big{\vert} ~\forall~ l,m \in i  \right\} \right) \label{Metrics_3}\
\end{align}
where $\text{Flow} \left( {{\rm{srv}}_l,{\rm{srv}}_m} \right)$ denotes the flow capacity between two of the available servers in the $i^{th}$ connected components. In the experimental results, we will study the $\text{Flow}_{\max}$ of the largest and second-largest pools. In the next section, the proposed analysis method that generates, evaluates, and aggregates snapshots to create the performance picture of a DC and its associated topology is discussed.

\hl{Another important factor to be considered in the performance metrics is the latency between nodes and also effective latency of workflows. We will address the latency performance in a future work in which a latency penalty is added to every component including servers and switches. The reason for considering explicit latency penalty for the servers and switches is the ongoing shift from application-specific integrated circuit (ASICs) to white-box and Linux-based software-defined components that makes their performance highly deponent on their actual stack of hardware/software inside the box.}

\section{Experimental Results and Discussions}
\label{sec_Res_Discussions}
In this section, the capability of the proposed Monte-Carlo approach is provided in the form of calculating the performance metrics of each topology against its parameters. This analysis is provided in the following three subsections.

\subsection{BCube topology}
\label{ssec_res-BCube}
\begin{figure}[tbh!]
\centering
\setlength{\tabcolsep}{1pt}
\setlength{\extrarowheight}{-5pt} \begin{tabular}{@{}cccc@{}}
(a) & \includegraphics[width=\figuresintablesix]{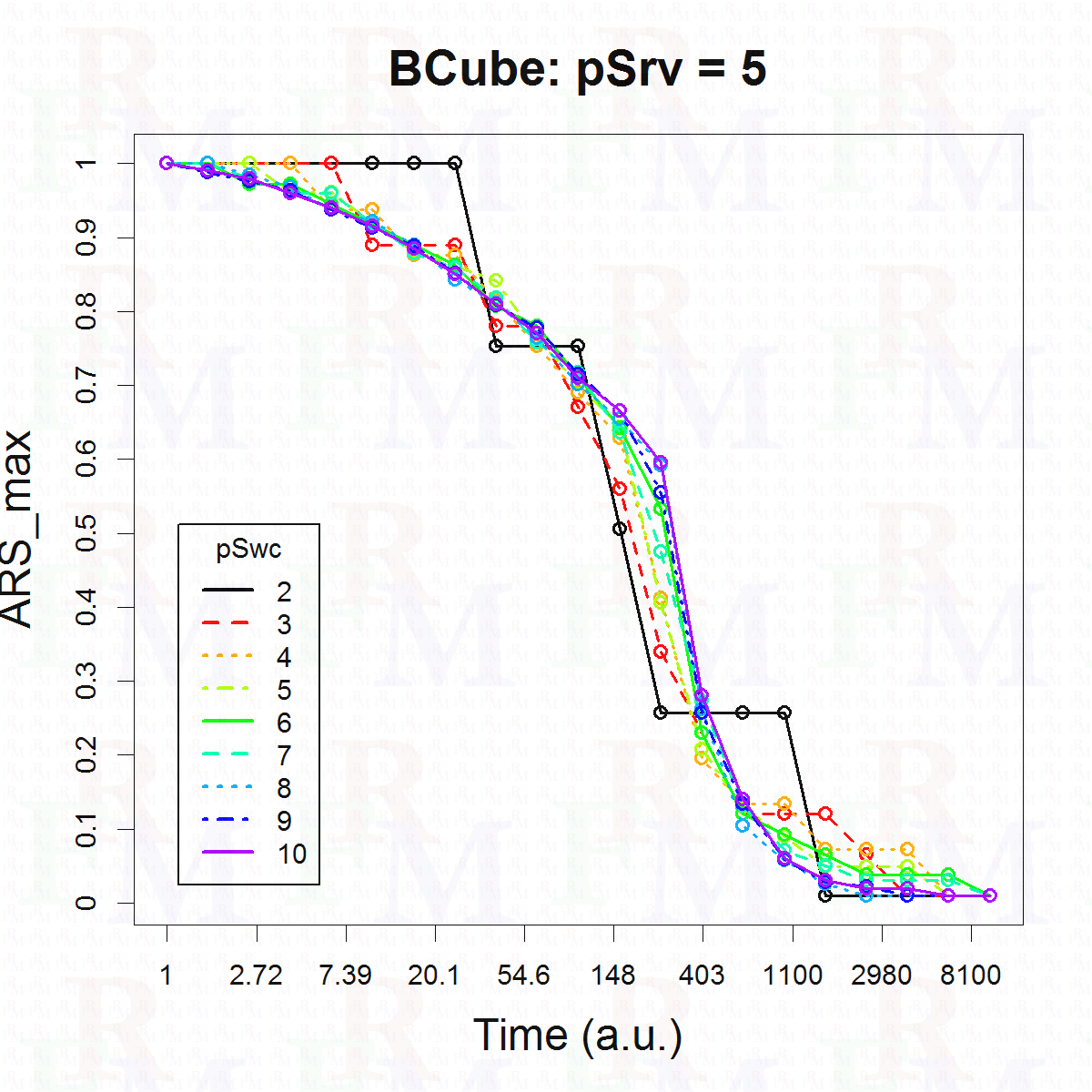} &
(b) & \includegraphics[width=\figuresintablesix]{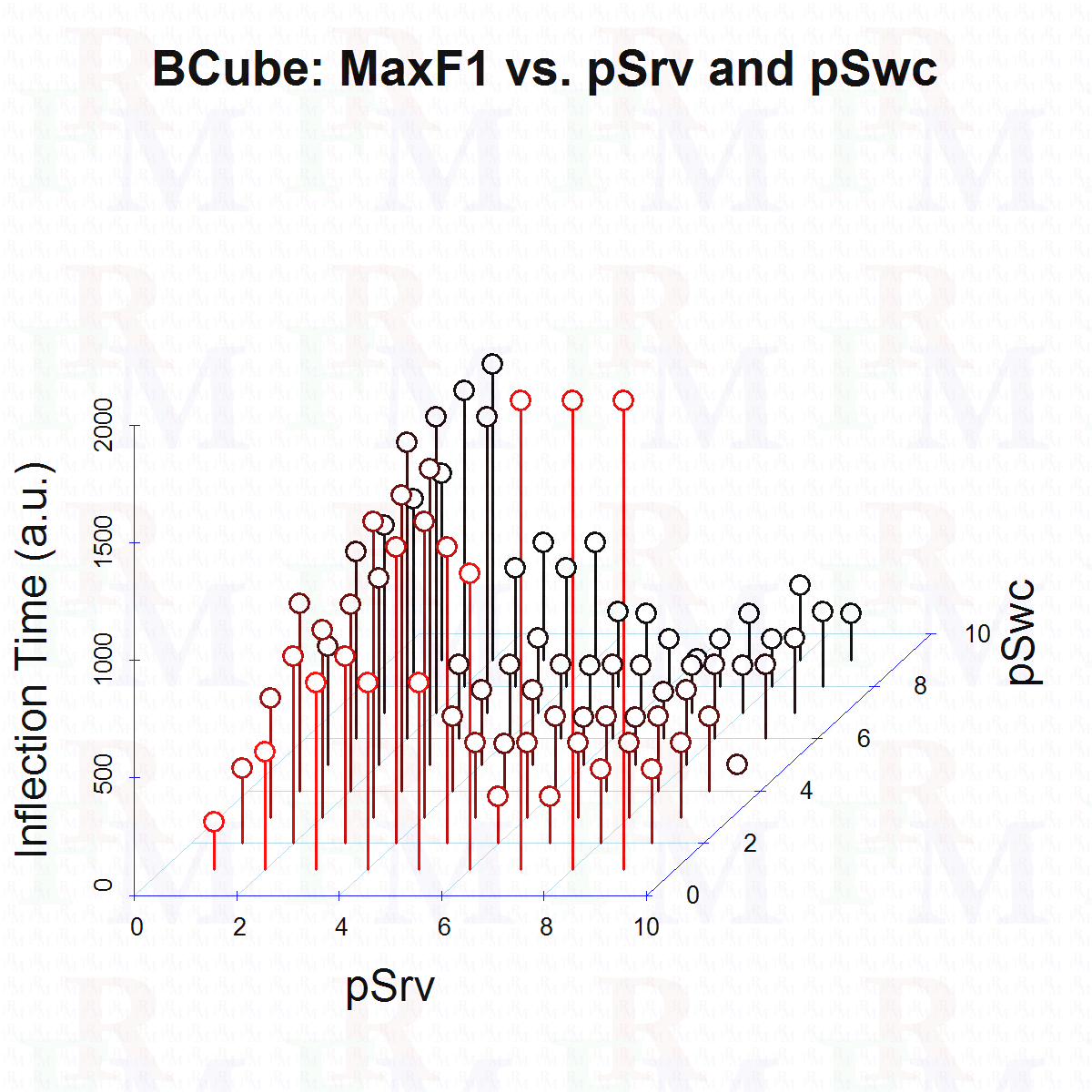} \\
(c) & \includegraphics[width=\figuresintablesix]{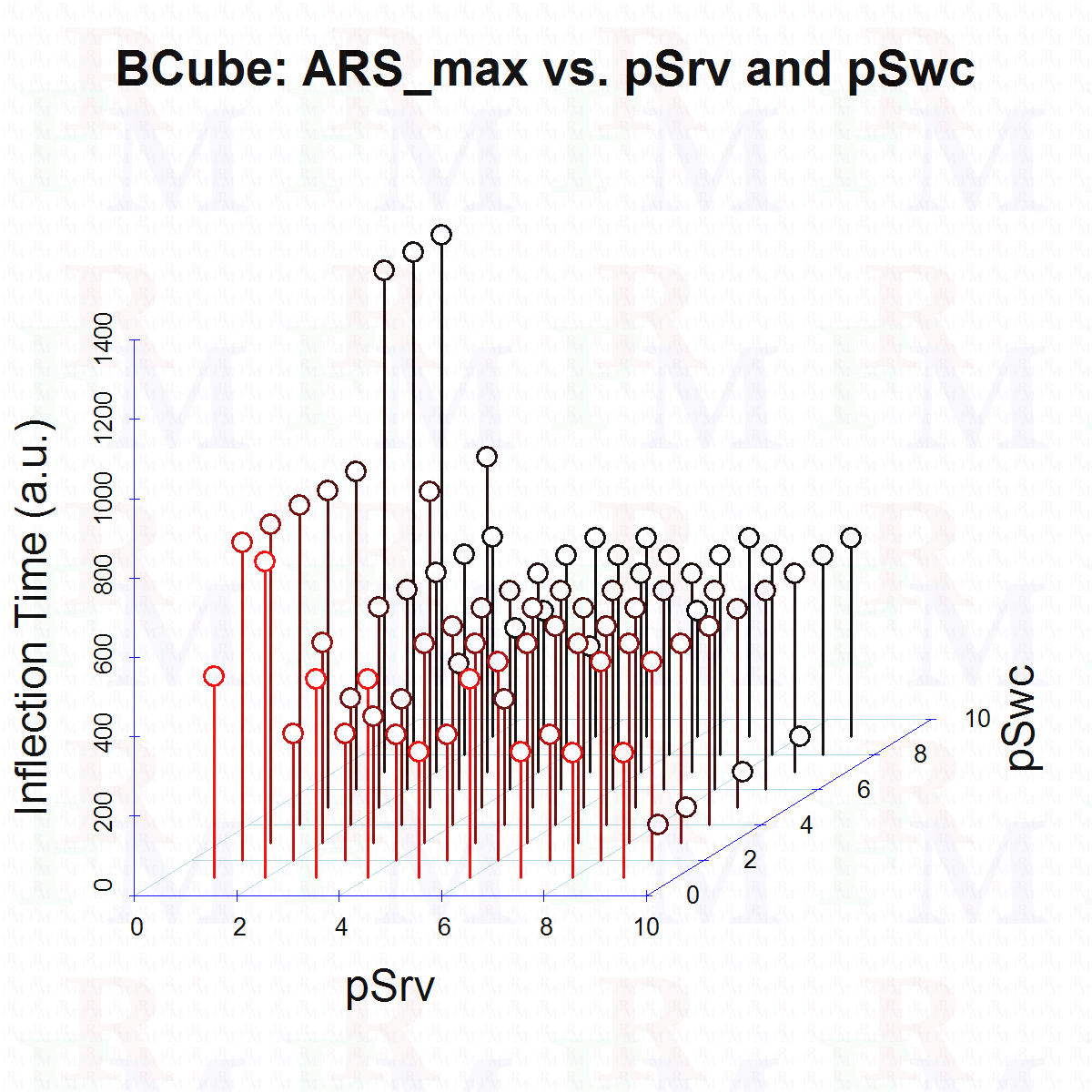} &
(d) & \includegraphics[width=\figuresintablesix]{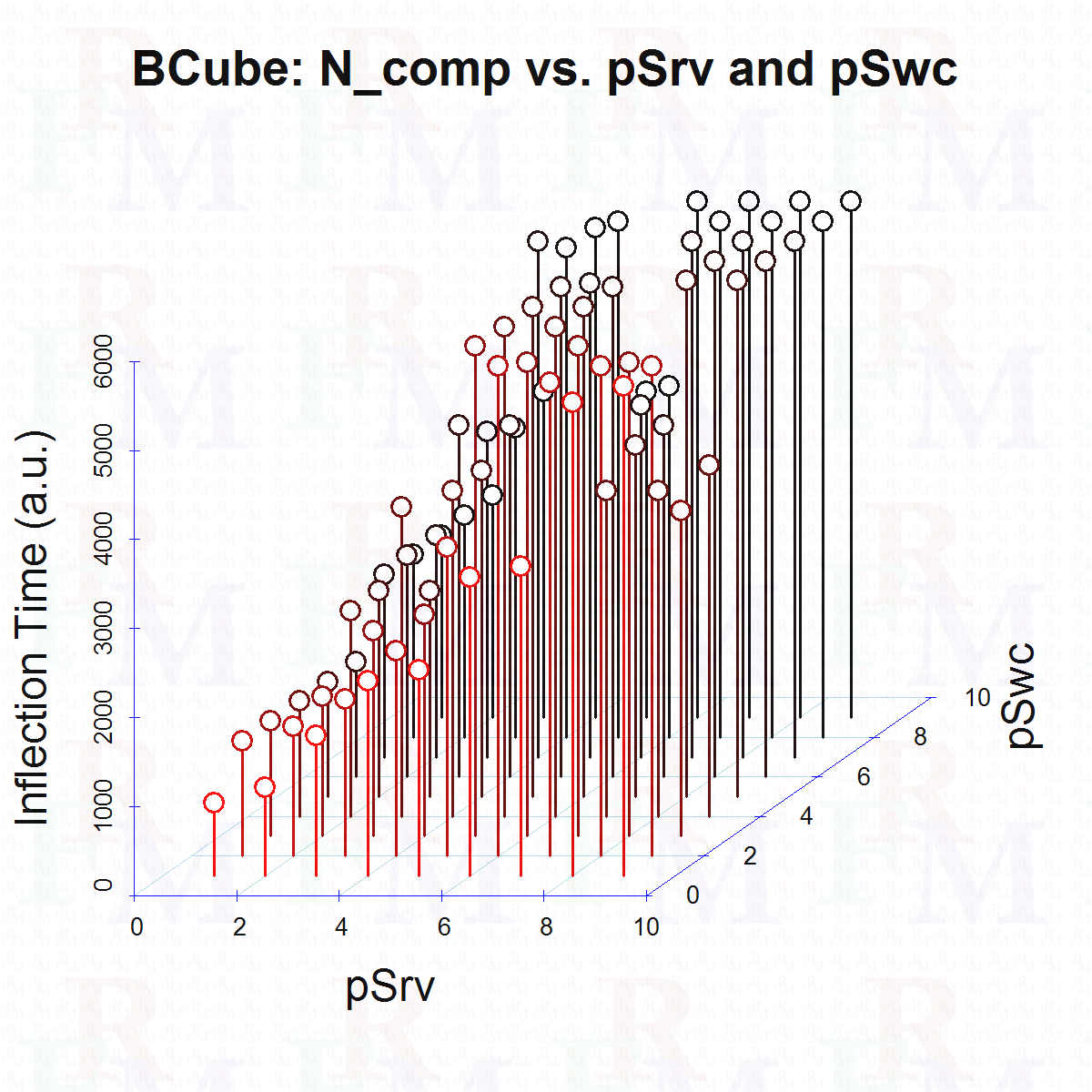} 
\end{tabular}
\caption{a) The profile of the $\text{ARS}_{\max}$ metric for the BCube topology.
b)-d) The variation of $\text{Flow}_{\max}$, $\text{ARS}_{\max}$, and $\text{N}_\text{comp}$ metrics versus the BCube topology's parameters.}
\label{f-Exp-BCube-against-parameters-1}
\end{figure}

As mentioned in section \ref{ssec_MDC-BCube}, the BCube topology has two parameters, $N=\text{pSrv}$ and $K=\text{pSwc}$. By definition, $N$ is also the number of switches in each level of the BCube, and the number of levels, and $K$ is the number servers in each BCube (level-0), and the number of the BCube (level-0)s in the zeroth level. The performance of BCubes against these two parameters is provided in Figures \ref{f-Exp-BCube-against-parameters-1}. In Figure \ref{f-Exp-BCube-against-parameters-1}(a), the performance of the $\text{ARS}_{\max}$ against the number of switches' port is illustrated. The number of servers' port is set fixed to 5. As can be seen with the increase in the number of ports, the behavior converge to a smooth and saturated form. The gain achieved at the inflection time, i.e., the time the performance curve changes its curvature behavior,\footnote{The inflection time is calculated using the Extremum Distance Estimator (EDE) method \cite{Christopoulos2012}.}
is almost \%100 in this figure. In the rest of plots in Figure \ref{f-Exp-BCube-against-parameters-1}, the inflection time is only provided in order to avoid plotting many curves. As can be seen, most of the performance parameters have improved behavior when the number of ports is increased. A comparison with other topologies and with higher and more practical number of servers and switches is provided in subsection \ref{ssec_usecase_3072}.

\subsection{FatTree topology}
\label{ssec_res-FatTree}
\begin{figure}[tbh!]
\centering
\setlength{\tabcolsep}{1pt}
\setlength{\extrarowheight}{-5pt} \begin{tabular}{@{}cccc@{}}
(a) & \includegraphics[width=\figuresintablesix]{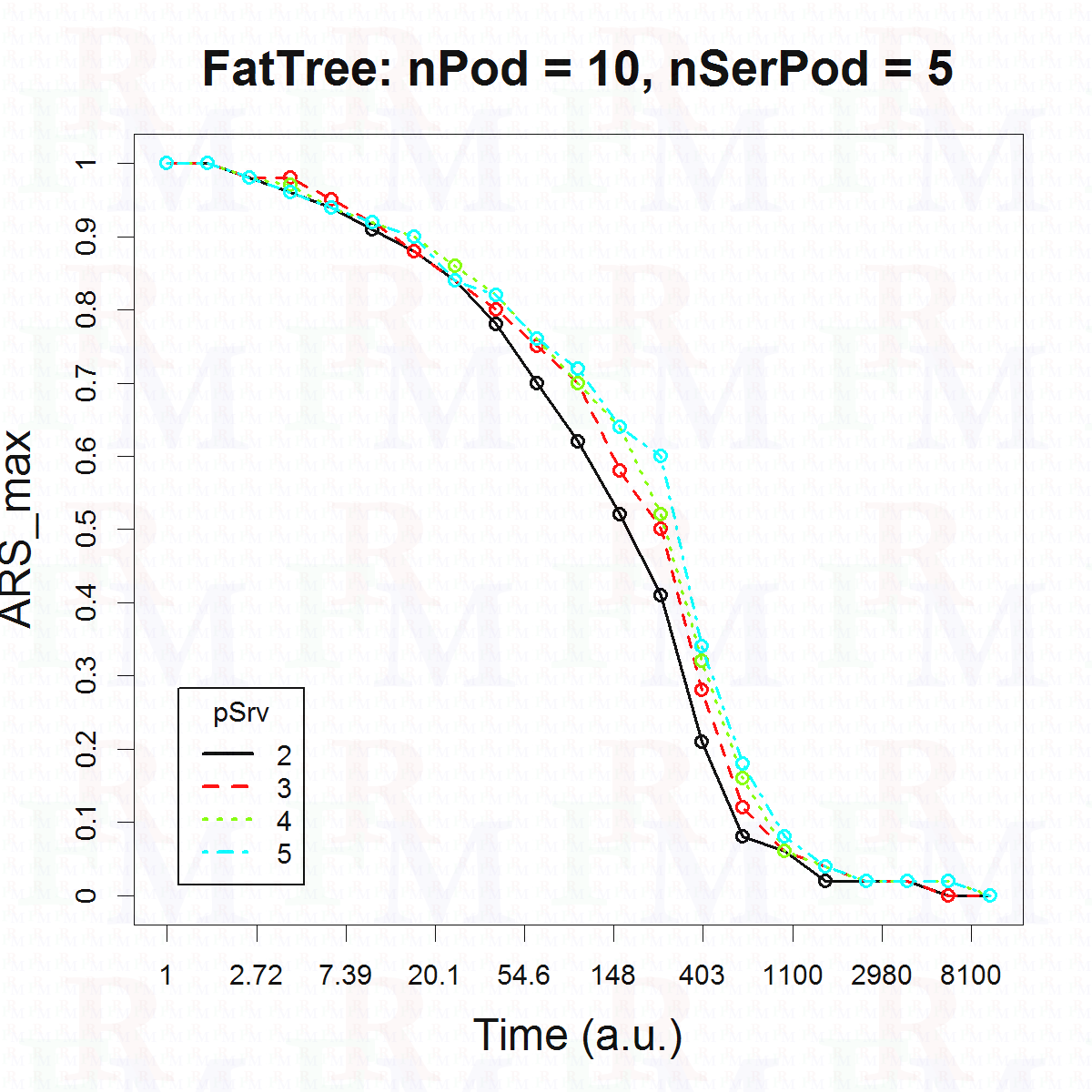} &
(b) & \includegraphics[width=\figuresintablesix]{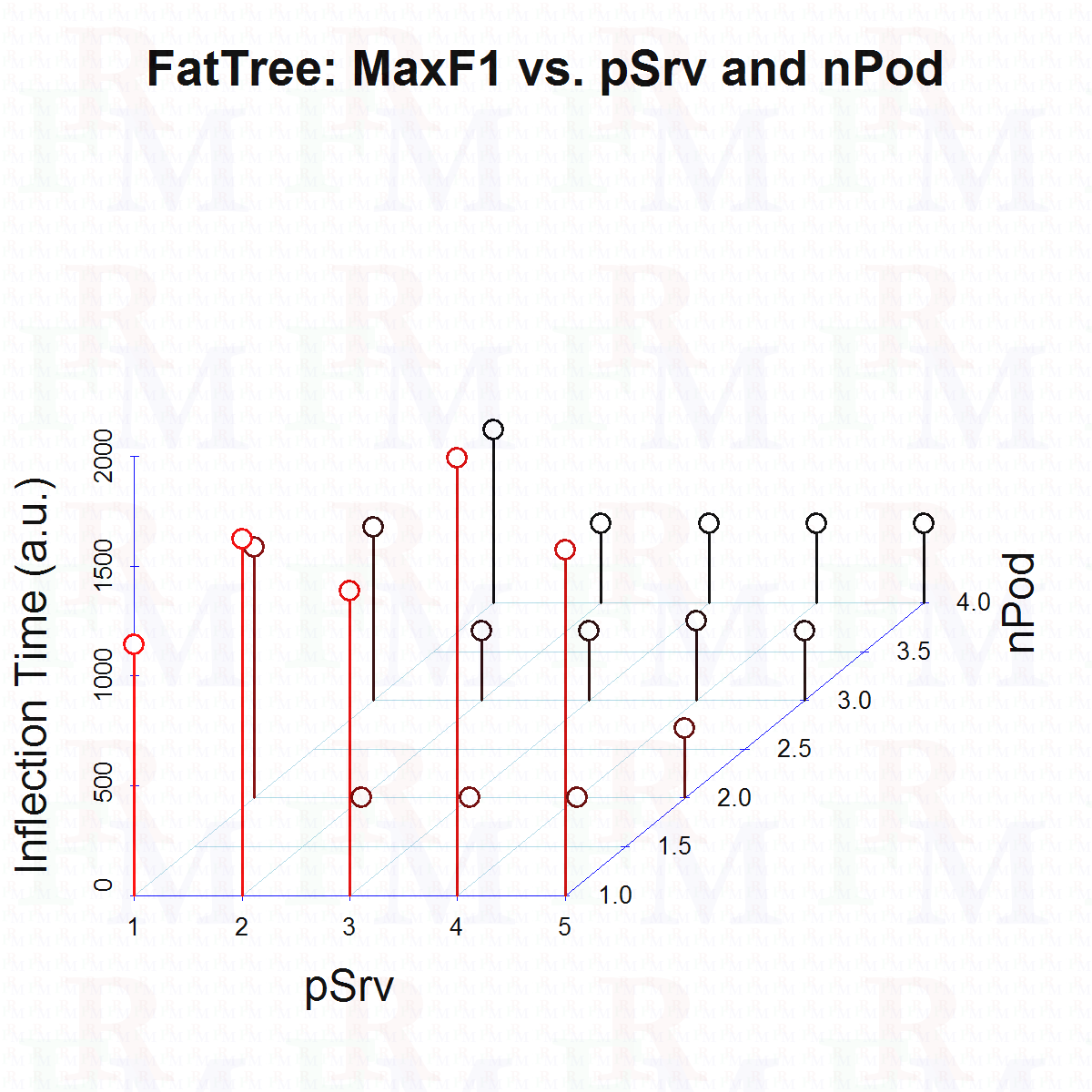} \\
(c) & \includegraphics[width=\figuresintablesix]{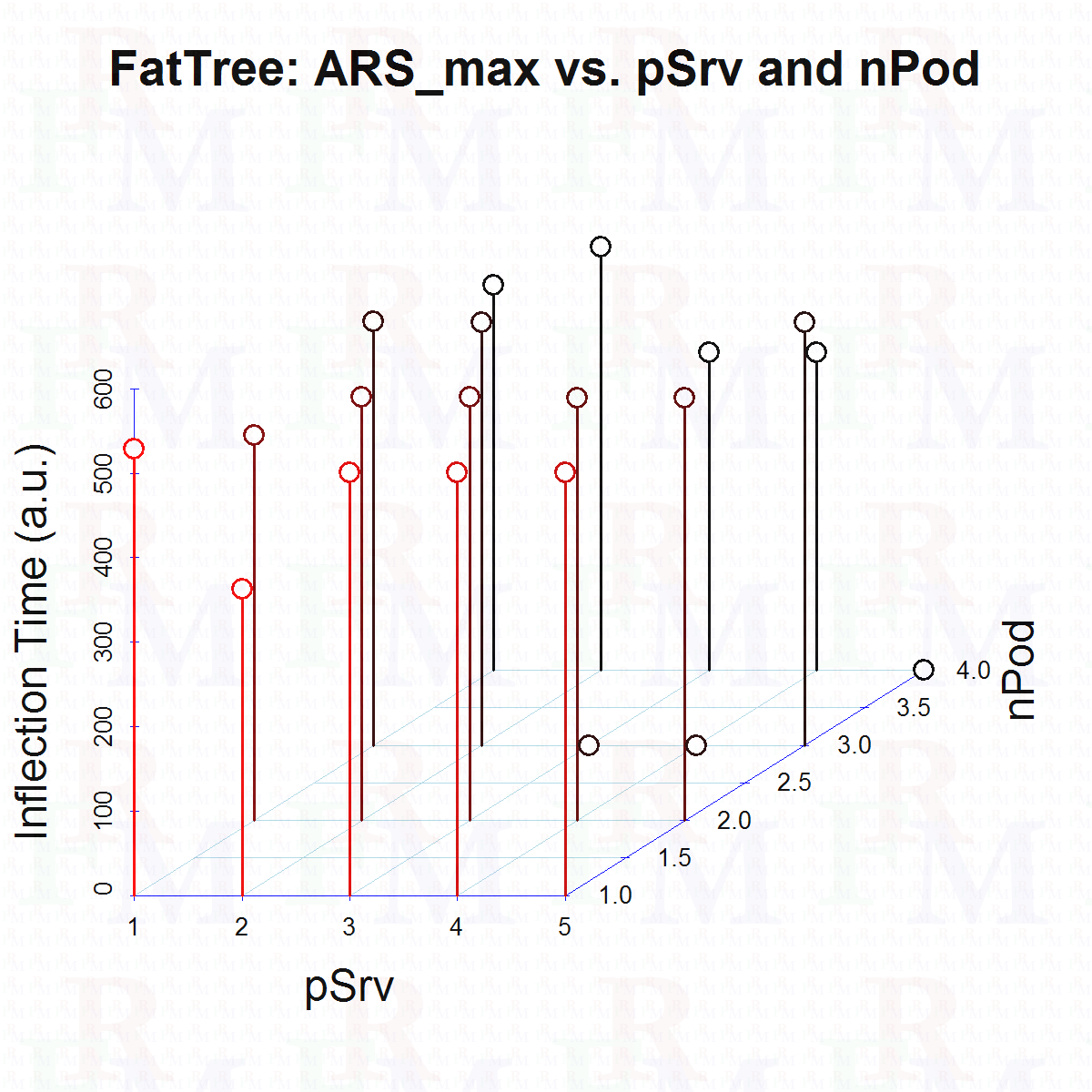} &
(d) & \includegraphics[width=\figuresintablesix]{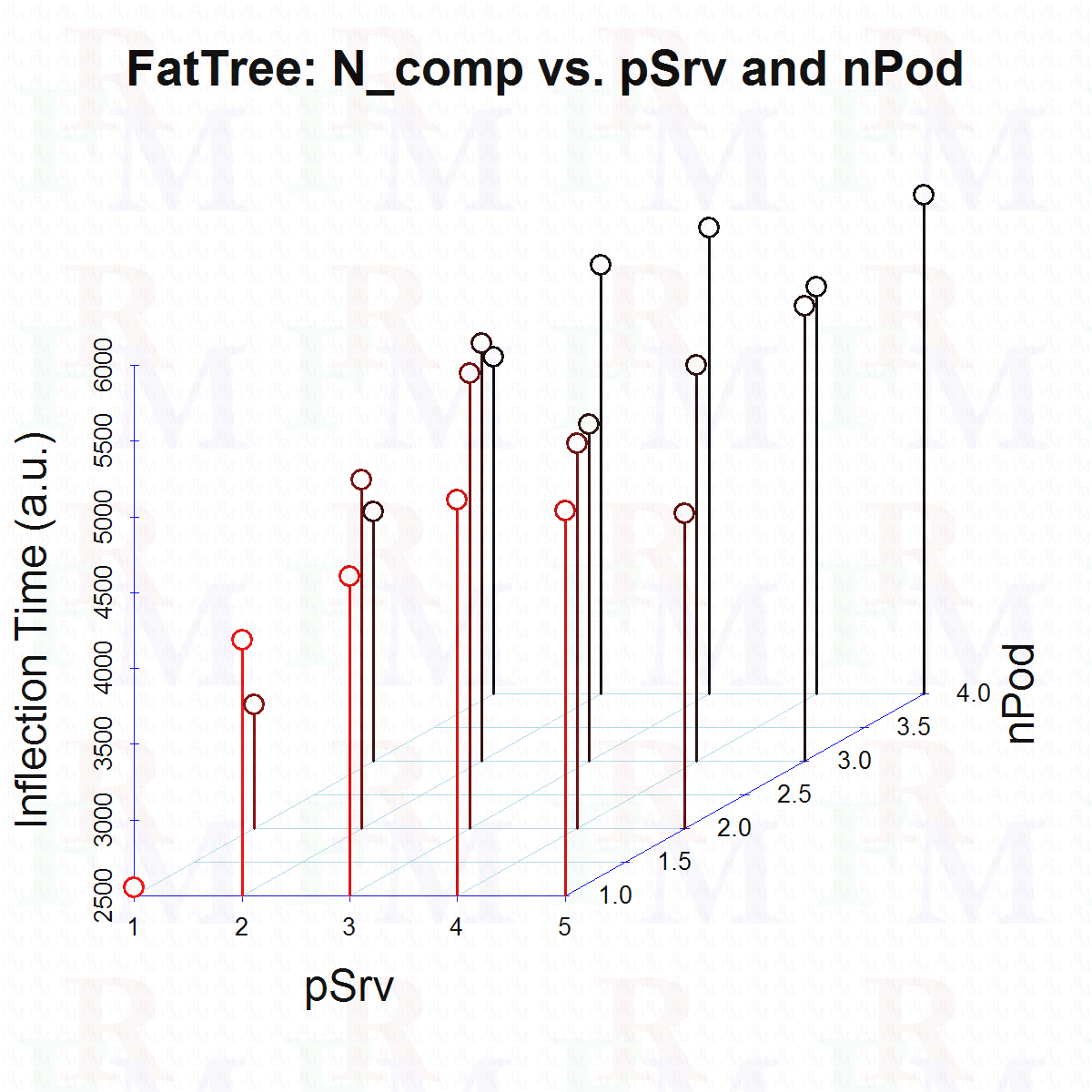} 
\end{tabular}
\caption{a) The profile of the $\text{ARS}_{\max}$ metric for the FatTree topology.
b)-d) The variation of $\text{Flow}_{\max}$, $\text{ARS}_{\max}$, and $\text{N}_\text{comp}$ metrics versus the FatTree topology's parameters.}
\label{f-Exp-FatTree-against-parameters-1}
\end{figure}

The FatTree topology is considered as the baseline in this work. The performance metrics behavior against parameters of this topology is provided in Figure \ref{f-Exp-FatTree-against-parameters-1}. The main parameters of FatTree topology, defined in section \ref{ssec_fattree_topo}, are the number of servers' port (pSrv), the number of pods (nPod), and the number of servers per pod (nSerPod). When comparing Figure \ref{f-Exp-FatTree-against-parameters-1}(a) and \ref{f-Exp-BCube-against-parameters-1}(a), it can be seen that the saturated performance of the FatTree topology is achieved by the BCube topology. The main cost associated with the BCube topology is bi-functional role of servers in this topology where they participate in data transmission among other nodes.

\subsection{MDCube2D topology}
\label{ssec_res-MDCube2D}
\begin{figure}[tbh!]
\centering
\setlength{\tabcolsep}{1pt}
\setlength{\extrarowheight}{-5pt} \begin{tabular}{@{}cccc@{}}
(a) & \includegraphics[width=\figuresintablesix]{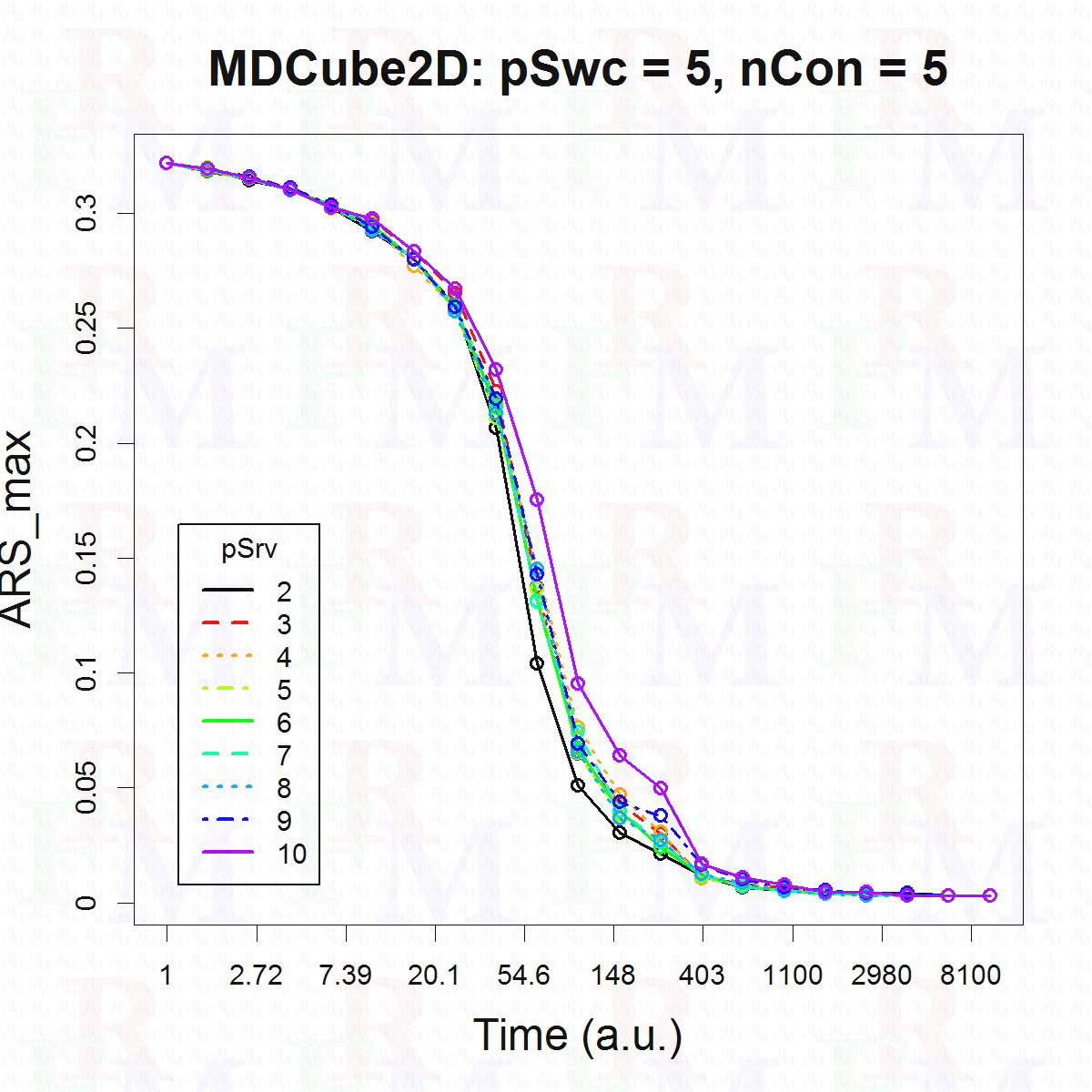} &
(b) & \includegraphics[width=\figuresintablesix]{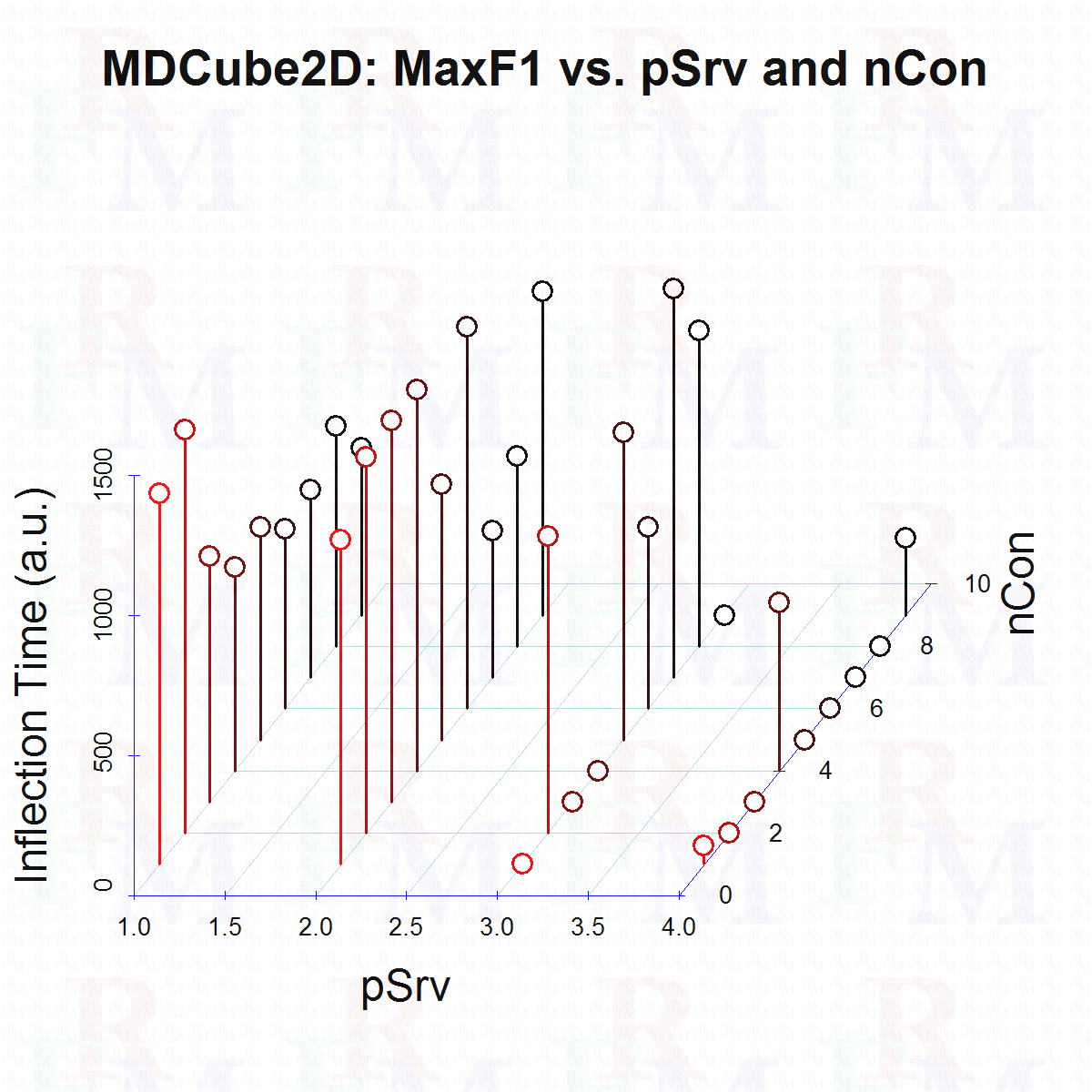} \\
(c) & \includegraphics[width=\figuresintablesix]{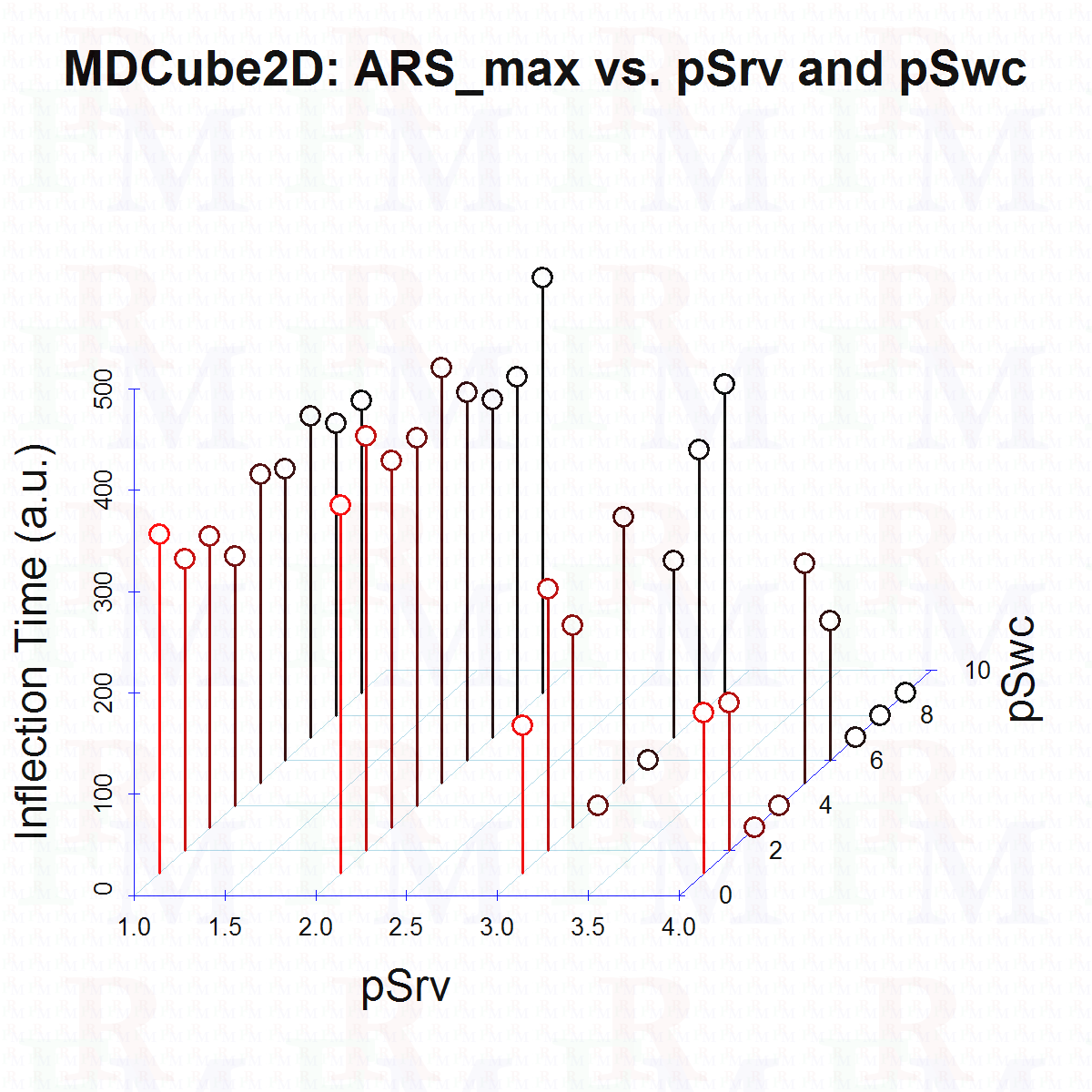} &
(d) & \includegraphics[width=\figuresintablesix]{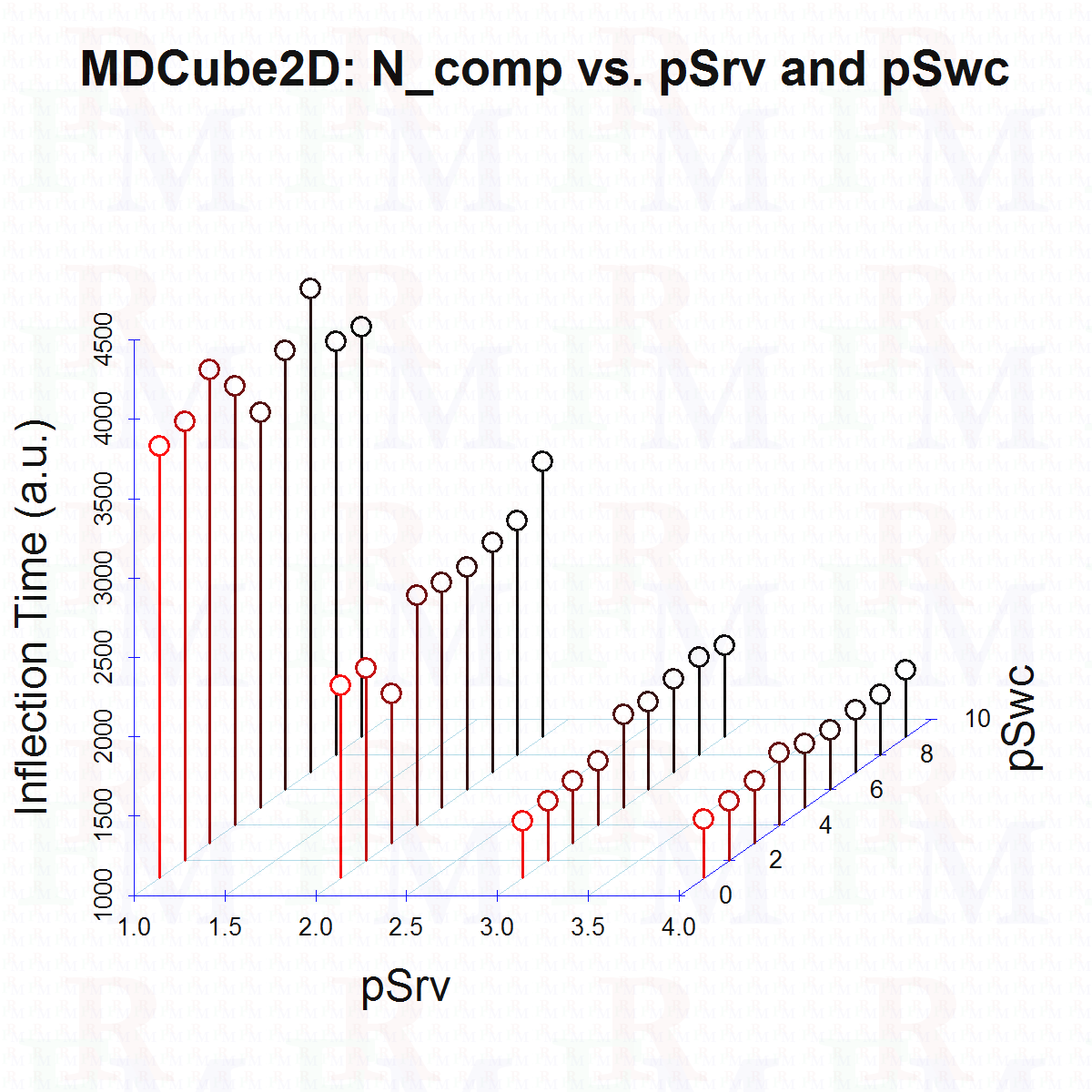} 
\end{tabular}
\caption{a) The profile of the $\text{ARS}_{\max}$ metric for the MDCube2D topology.
b)-d) The variation of $\text{Flow}_{\max}$, $\text{ARS}_{\max}$, and $\text{N}_\text{comp}$ metrics versus the MDCube2D topology's parameters.}
\label{f-Exp-MDCube2D-against-parameters-1}
\end{figure}

The same process of evaluating performance metrics against the parameters of the MDCube2D topology were carried out and is provided in Figure \ref{f-Exp-MDCube2D-against-parameters-1}. Obviously, the MDCube2D topology has less performance in terms of $\text{ARS}_{\max}$ as can be seen from Figure \ref{f-Exp-MDCube2D-against-parameters-1}(a). This can be associated with the bottlenecks created by the interconnecting links among the containers. However, the behavior of this topology is more stable across the parameter values, which can be seen from Figure \ref{f-Exp-MDCube2D-against-parameters-1}(c). The other advantage of the MDCube2D topology is its by-definition scalability. When the BCube topology's number of servers is increased, the number of level-0 cubes and also the complexity of the connections rapidly increase. In contrast, MDCube2D topology is highly scalable in a 2D manner, which fits well with the spatial organization of the traditional DC rooms. 

The compromise between scalability and performance can be decided on depending on the application context. For example, for a central data center, which requires adding new modules and hardware along the operating in order to address the needs of new customers, it can be suggested that the MDCube2D topology is a better choice in terms of scalability. In contrast, for example, in a small edge data center used by a Telco operator to service its customers in a small area, the BCube topology seems to be the better choice. In the next section, section \ref{ssec_usecase_3072}, we compare the performance of the extended topologies proposed in this paper against each other in a practical scale of a few thousands of servers. 

\subsection{A Use Case of 3072-server MDC: An Application of BCube$^{E}$, MDCube2D$^{E}$, and FatTree$^{E}$ Architectures}
\label{ssec_usecase_3072}
All three considered topologies show a considerable resilience to failure of their components along time. However, as observed in the previous sections, the degree of these residencies varies among the topologies. Also, the comparison at the scales similar to real word DCs should be performed. In this section, the performance comparison is performed for the same number of 3,072 servers in each topologies. 
Using 48-port switches, a 2304-server BCube (level-1) architecture can be achieved. The next DC size would be 2704 using 52-port switches. Similarly, the FatTree topology could achieve the same number of servers using these switches. In contrast, our proposed BCube$^{E}$ topology can provide an level-1 architecture using 64 BCube (level-0) cubes each one containing 48 servers to easily archive the target 3072-server size.

 A comparison table of the required number of servers and switches of the three extended topologies is provided in Table \ref{t-bcubes-param1}. In particular, the MDCube2D$^{E}$ architecture composed of four  BCube$^{E}$ architectures each containing 768 servers. The FatTree$^{E}$ architecture composed of 64 pods, and the BCube$^{E}$ architecture has 64 level-0 cubes.

\begin{table}
\centering
\small
\begin{tabular}{||c||c|c|c|c||}\hline\hline
Topology		& pSrv	& pSwc	& \# servers& \# switches	\\\hline\hline
BCube$^{E}$			& 2		& 48	& 3072		& 128			\\\hline
FatTree$^{E}$		& 2		& 48	& 3072		& 134			\\\hline
MDCube2D$^{E}$		& 2		& 48	& 3072		& 128			\\\hline\hline
\end{tabular}
\caption{The parameters and specifications of the three MDC topologies used for the 3072-server use case.}
\label{t-bcubes-param1}
\end{table}

\begin{figure}[tbh!]
	\centering
	\setlength{\tabcolsep}{1pt}
	\setlength{\extrarowheight}{-5pt} 	\begin{tabular}{@{}c@{}@{}c@{}@{}c@{}@{}c@{}}
		(a) & \includegraphics[width=\figuresintablefour]{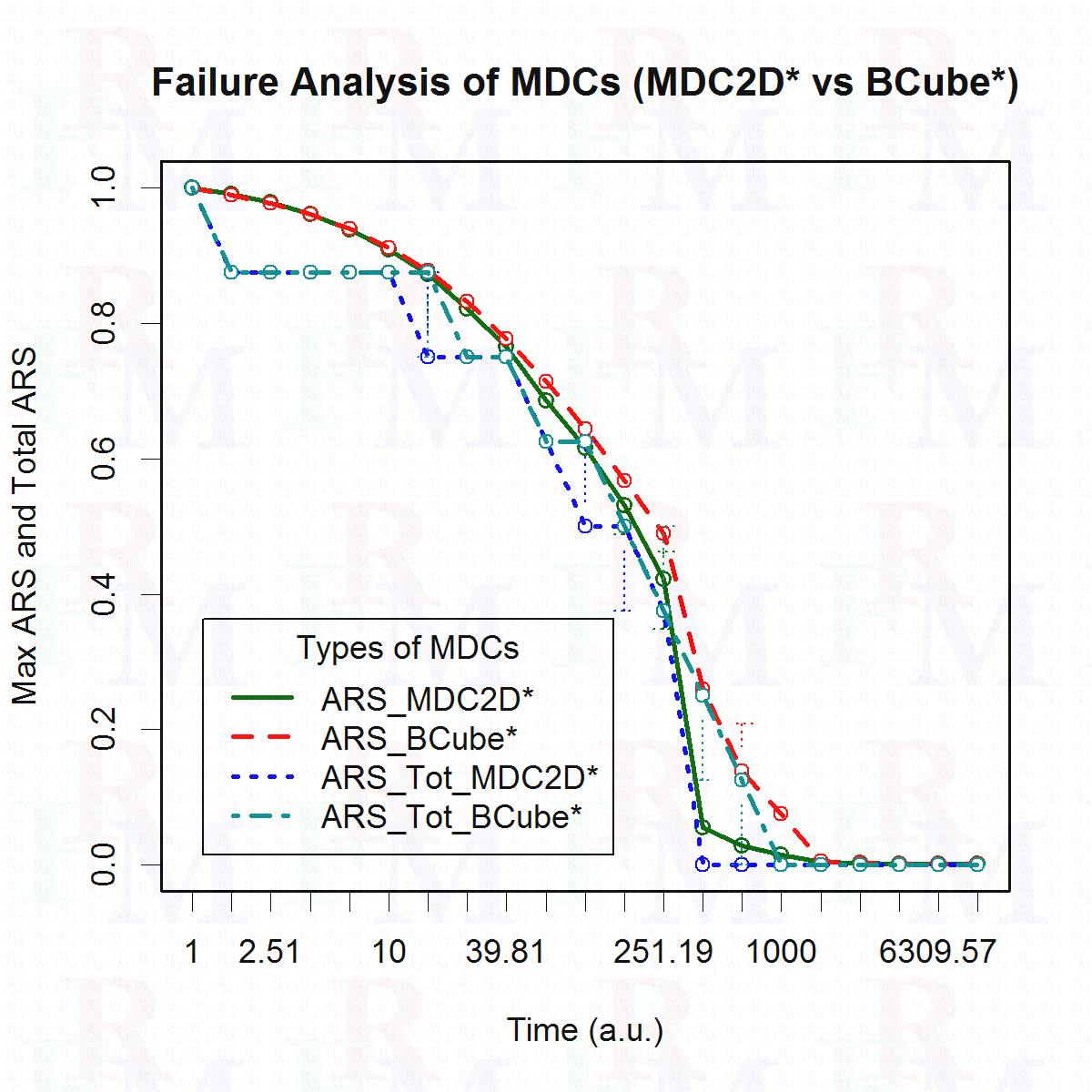} &
		(b) & \includegraphics[width=\figuresintablefour]{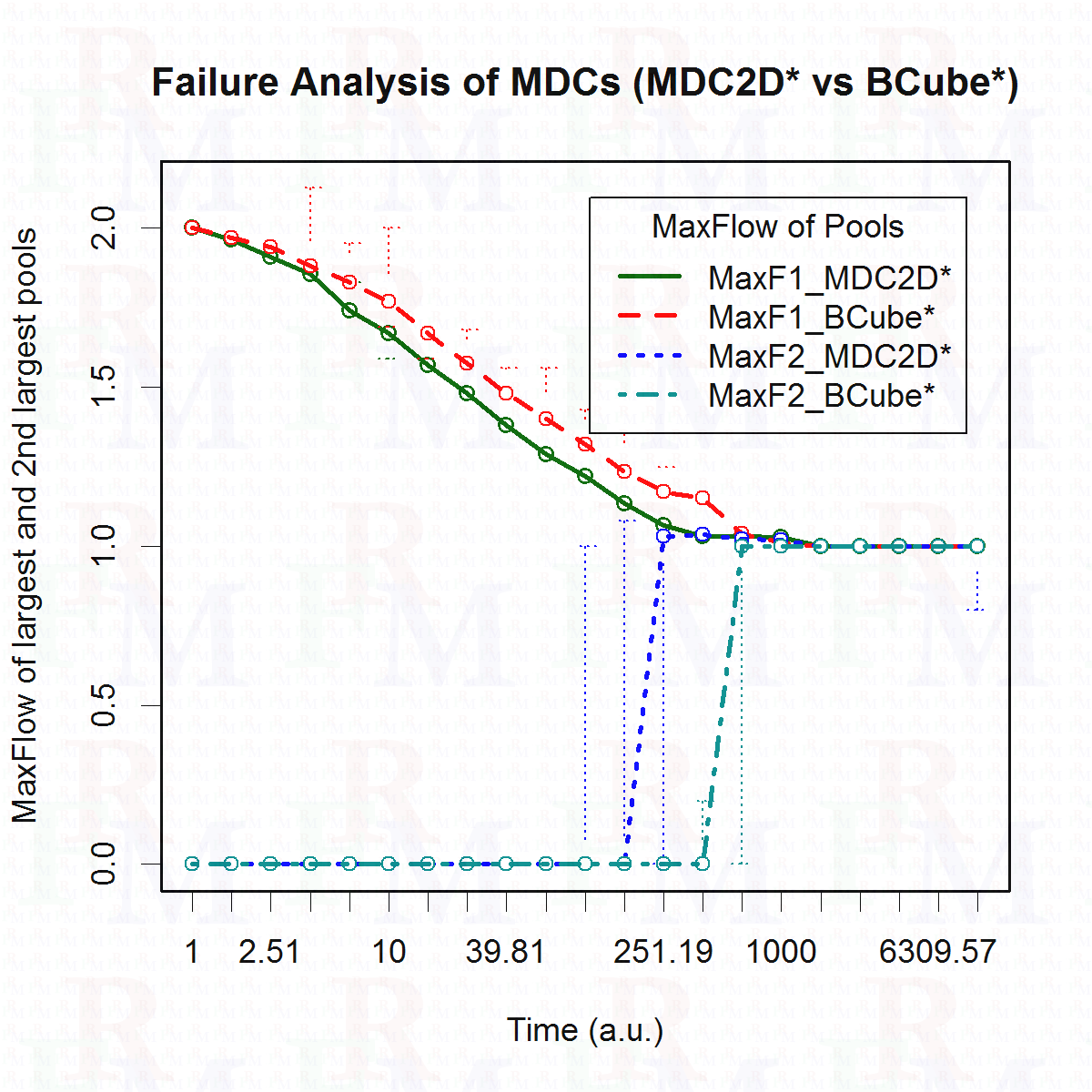} \\
		(c) & \includegraphics[width=\figuresintablefour]{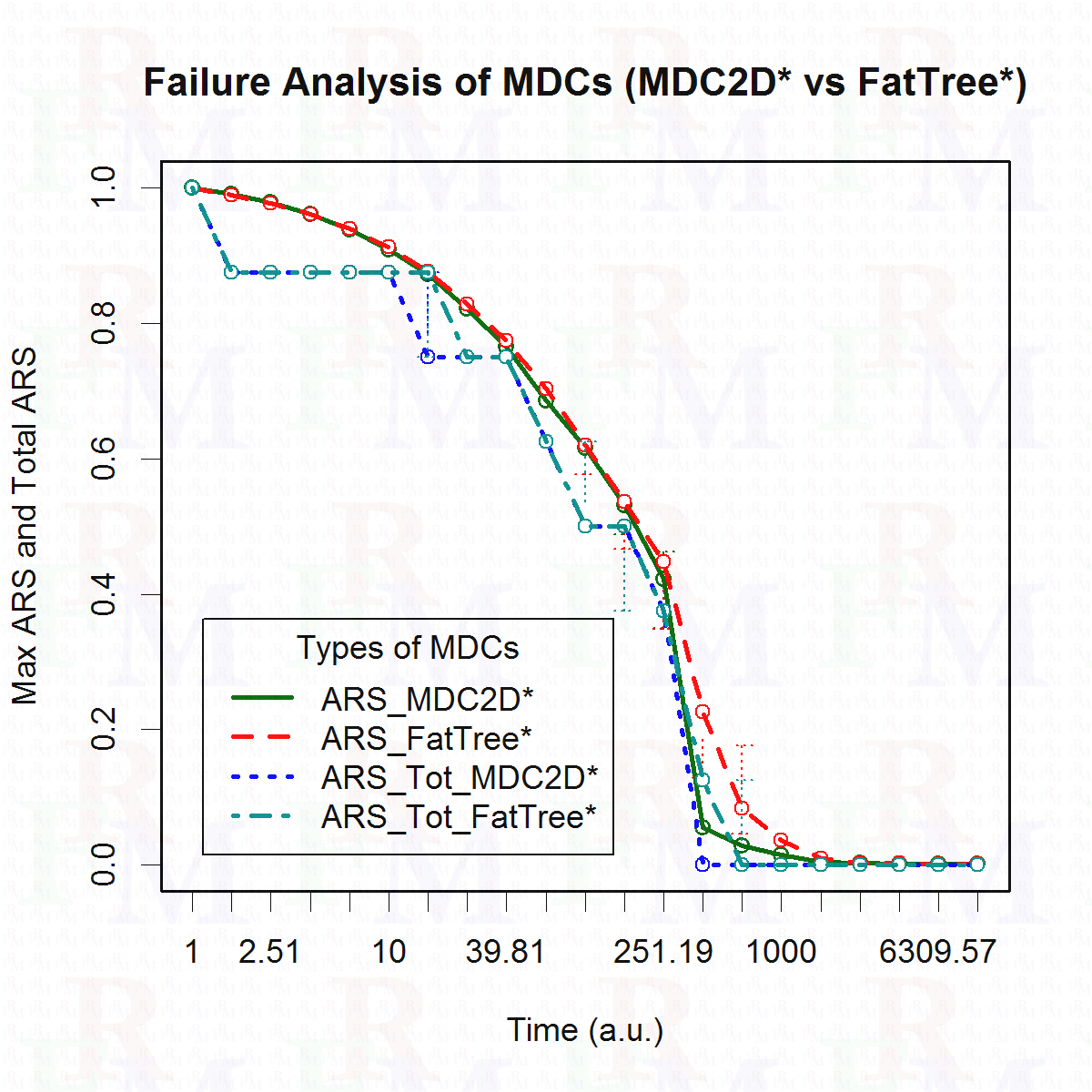} &
		(d) & \includegraphics[width=\figuresintablefour]{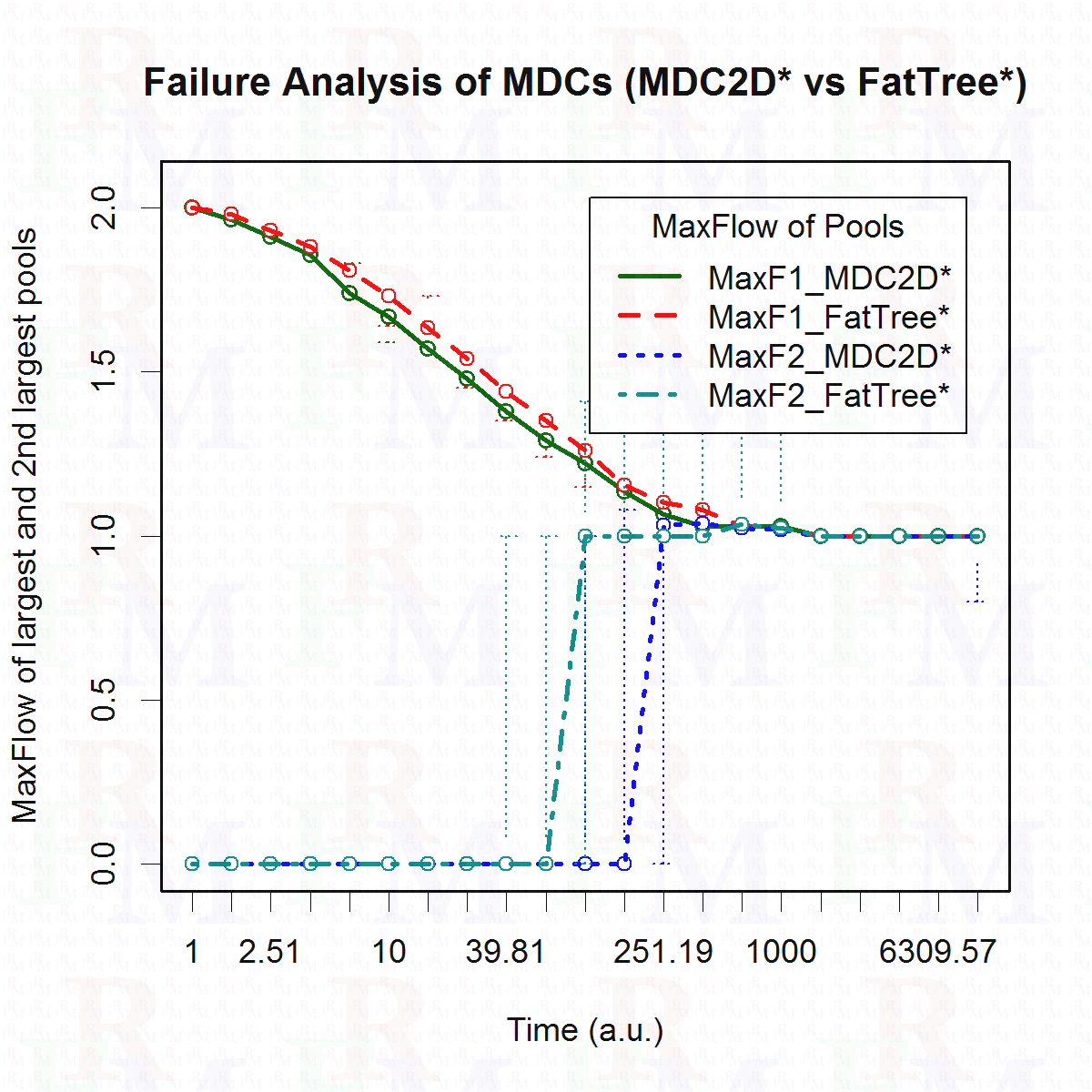} \\
		(e) & \includegraphics[width=\figuresintablefour]{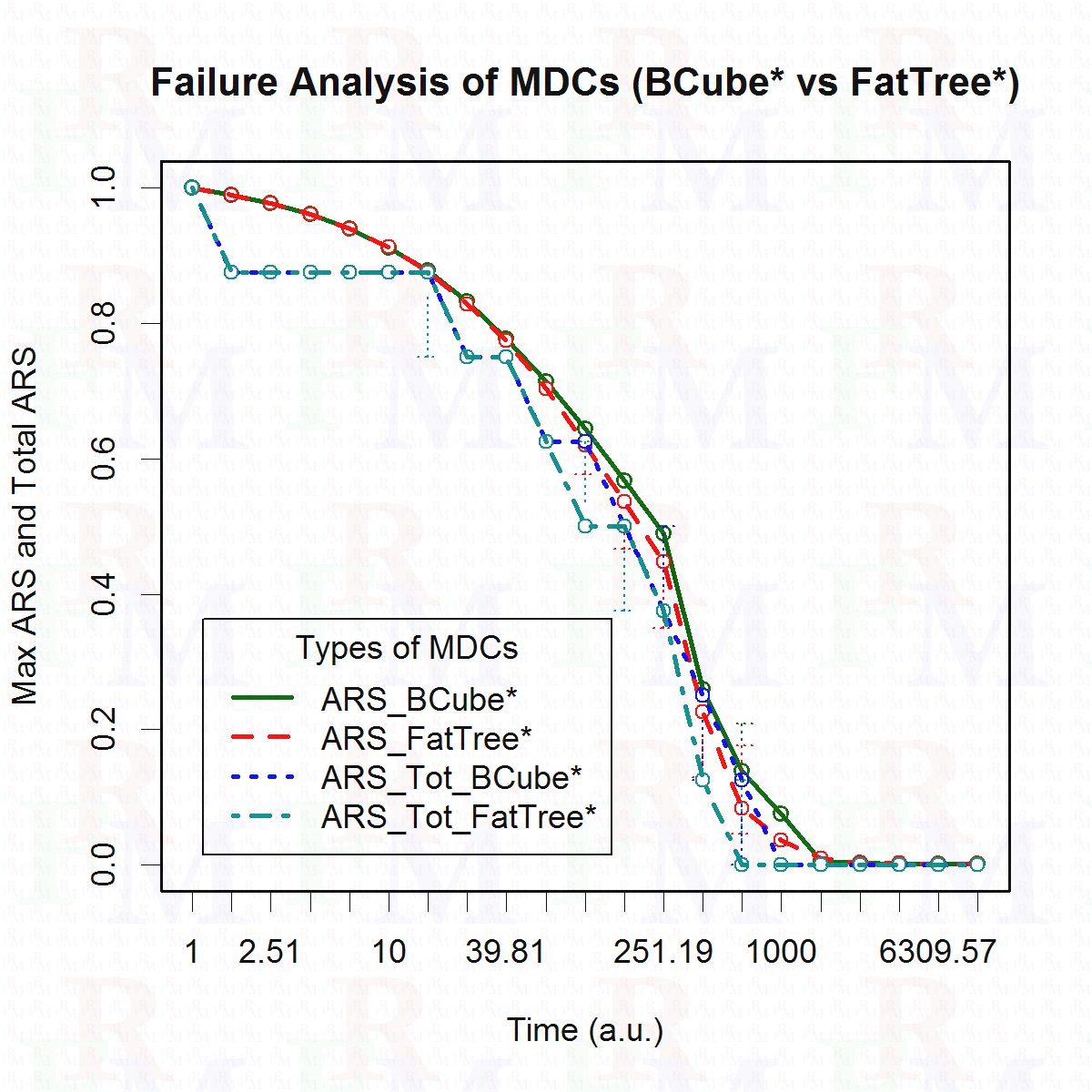} &
		(f) & \includegraphics[width=\figuresintablefour]{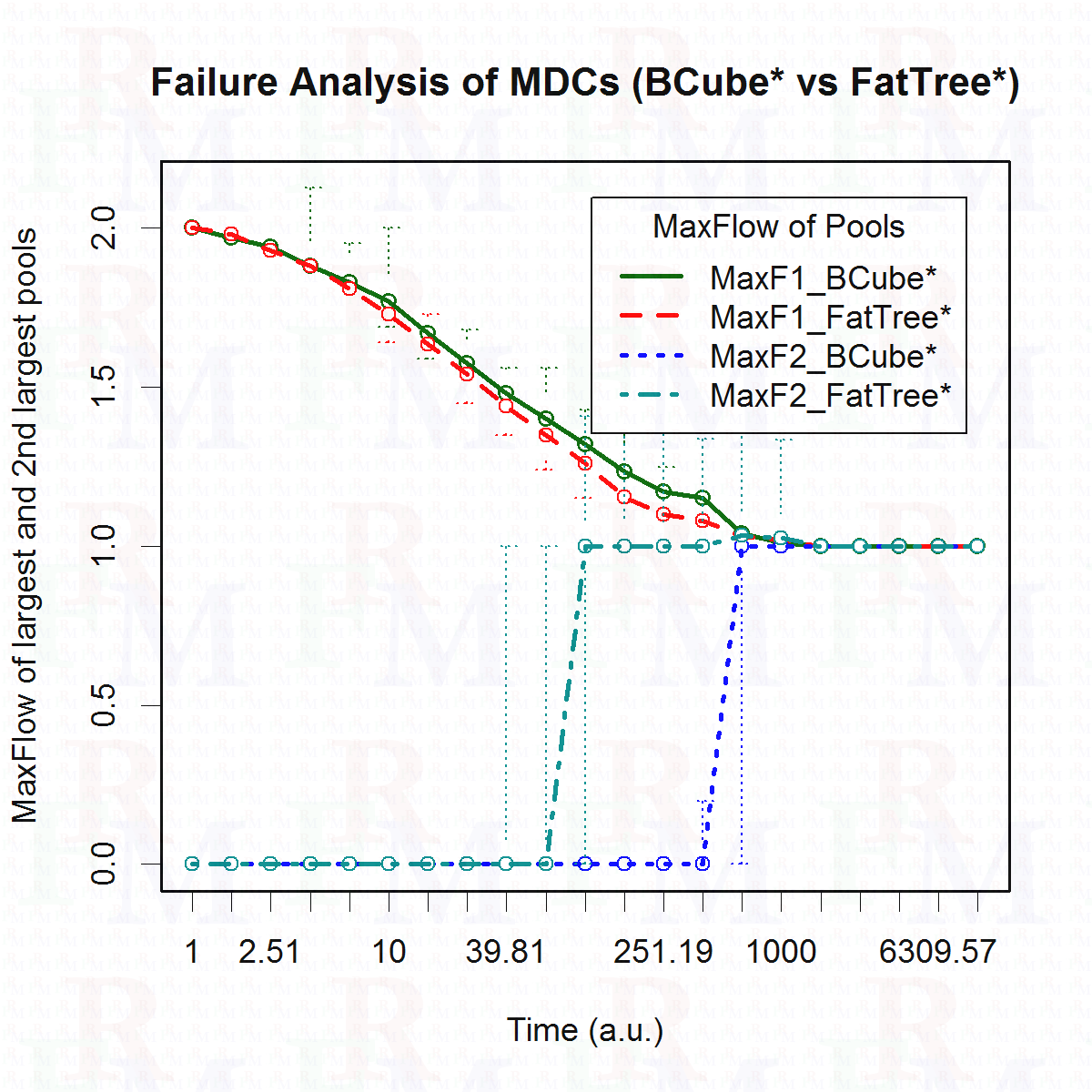} 
	\end{tabular}
	\caption{The pairwise comparison of the three proposed topologies using the $\text{ARS}_\text{max}$ and $\text{Flow}_\text{max}$ metrics. 
		a)-b) MDCube2D$^{E}$ vs BCube$^{E}$.
		c)-d) MDCube2D$^{E}$ vs FatTree$^{E}$.
		e)-f) BCube$^{E}$ vs FatTree$^{E}$.}
	\label{f-Exp-MDC2DS-vs-BCubeS-vs-FatTreeS-1}
\end{figure}

First, the comparative performance of the three designs is analyzed in Figure \ref{f-Exp-MDC2DS-vs-BCubeS-vs-FatTreeS-1}. Five metrics 
are considered. The $\text{ARS}_\text{Tot}$ closely follows the $\text{ARS}_{\max}$ that is a good performance for the BCube topologies in profitability of the solution by providing service to multiple customer or tenants along the time. The resilience to breakdown to multiple pools can be seen from the $\text{MaxF}_{1}$ and $\text{MaxF}_{2}$ performance. Considering the 2-port servers included in the design, the upper limit of the $\text{MaxF}_{1}$ is 2, and it degrades along the time toward 1. The $\text{MaxF}_{1}$ does not go below one because of its definition that calculates the flow within a pool and island. Interestingly, the breakdown time period of the FatTree$^{E}$ architecture does not match the period of degradation associated with the biggest pool. The breakdown is around time equal to 80, while the biggest pool degrades to a flow of one around time equal to 700. This can be explained by multiple partitioning events that happen at the breakdown time and result in several small pools. In contrast, the BCube$^{E}$ and the MDCube2D$^{E}$ topologies, observe a partitioning into a big and small pool at the breakdown. 

The BCube$^{E}$ topology performs the best in terms of three measure: $\text{ARS}_{\max}$, $\text{ARS}_\text{Tot}$, and $\text{MaxF}_{1}$. On the second rand, the MDCube2D$^{E}$ topology stands with an exception with respect to the $\text{MaxF}_{1}$. The bottleneck of the MDCube2D$^{E}$ topology can be spotted in its inter-container links that can easily downgrade its overall flow performance.

\hl{It is worth mentioning that the default use case of the proposed analysis is in the design stage in order to select or determine the optimal network topology for the data center. However, the proposed analysis could be used as a ``live'' optimization tool in an in-operation data center in order to locally optimize the current ``brown'' topology. By brown topology we mean having components aged and even some failed in the system. This live optimization could help the data center operator to postpone the maintenance events as far as possible which is a key factor in operation of modular data centers that are deployed in remote areas.}

\hl{In terms of the overhead of analysis in the live-optimization use cases, it is worth mentioning that the snapshot concept itself is highly parallelizable, and therefore the required time to perform the live analysis could be arbitrary reduced by increasing the amount of its allocated parallel computation.}

\section{Related Work}
\label{sec_Related_Work}

There are several work addressing the reliability concern in DC networks \cite{Guo2009,Guo2008,Popa2010,Shang2012,Couto2012}. In these studies, a few DC network architectures were investigated in order to provide a better reliability and scalability with MDCs. 
In this regard, a comparison study on the reliability of DC topologies such as FatTree, BCube and DCell, were investigated in \cite{Guo2009,Guo2008} subject to a specific traffic pattern and routing algorithm, respectively. In particular, it has been shown that BCube topology that is specifically designed for MDCs using shipping-containers, shows a better performance than FatTree and DCell topologies for MDCs.  
Moreover, there are number of studies those taken into account comparisons based on the energy conservation and cost of different DC network architectures \cite{Popa2010,Shang2012}. For instance, the authors in \cite{Popa2010} have taken one step to understand the trade-offs in the design of DC network architectures. In particular, they proposed a methodology to estimate the cost of various DC architectures based on a predetermined performance requirement.

Recently, independent of any network applications, the authors in \cite{Couto2012} investigated the reliability of different DC network topologies in terms of the network components' failure ratios. In these analysis, the DC network topologies were modeled as a graph with servers and switches as nodes and network links that connecting the nodes. 
The results has shown a degradation at the DC network performance because of the removal of connected components with a small number of servers as the failure ratio increases. 
However, please note that the reliability comparisons analysis provided in studies like \cite{Couto2012} were utilized the removal of different network components to evaluate the effect of each component's failure on the reliability of different DC network topologies. 
It is worth noting that though such a comparison is common and used in many contributions, it is also challengeable since it does not consider the failure  behavior of each component over time. In fact, up to our knowledge, there exists no study on comparing different DC network topologies in terms of failures on DC network components while taking into account the operational life-time (lifespan) of the network.

\afterpage{\clearpage}

\section{Conclusion and future prospects}
\label{sec_conclusion}
In this work, a novel statistical approach, based on the Monte-Carlo methodology, has been proposed to estimate the performance metrics of modular data centers along their operating lifespan. To our knowledge, this is the first time such an analysis is provided. The proposed approach breaks down the trajectory dependency of the system in order to develop an unbiased statistical picture of a design's performance along its utilization. The approach uses the components probability distributions of failure along their utilization time in order to calculate the failure penetration at the components level, referred to as a snapshot. At the same time, to have a better fit on the real system failure data, a generalization in the form of a Tanh-Log cumulative probability distribution has been proposed. Using the proposed distribution and the analysis approach, the performance of three well-known topologies in the context of modular data centers, i.e., FatTree, BCube, and MDCube2D, has been studied. In addition, in order to make these topologies more flexible and agnostic with respect to hardware, and increase their resilience to failures, some extended versions of these topologies have been introduced and denoted as FatTree$^\text{E}$, BCube$^\text{E}$, and MDCube2D$^\text{E}$. It has been concluded that the extended BCube topology, BCube$^\text{E}$, could deliver more resiliency performance in terms of various performance metrics (subject to the validity range of the distributions used). On the other end, with respect to scalability, the extended MDCube2D, i.e., MDCube2D$^\text{E}$, provides a higher performance. 

Study of the impact of other layers and components, such as virtual machine managers, virtual machines, containers, and in general the software part on the reliability, availability, and management of the modular data centers will be considered in the future. Also, possible advantages of the Tanh-Log distribution on various dataset will be examined. Moreover, other potentials of the proposed extended topologies and also their possible drawbacks will be studied. 

\setstretch{1.0}
\small

\section*{Acknowledgments}
The authors thank the NSERC of Canada for their financial support under Grant CRDPJ 424371-11 and under the Canada Research Chair in Sustainable Smart Eco-Cloud (NSERC-950-229052), and also the MITACS of Canada. They also thank the anonymous reviewers for the insightful comments.

\section*{References}

\bibliography{imagep}   

\appendix

\setcounter{table}{0}
\setcounter{figure}{0}

\section{Tanh-Log and c-Tanh-Log Failure Probability Distributions}
\label{sec_Tanh}
In our approach to failure analysis, each component is assumed to follow a specific type of failure profile specific for its type of components. Also, we assumed that the system would be under full utilization, and therefore utilization time (age) of every surviving component is equal to each other 
This implicitly implies that the system has just one cold start and there will no (hardware) repair in the life-time of the system before major maintenance cycles. This is compatible with our MDC use cases in which the systems are shipped, deployed, and remotely controlled until their RoI goes beyond a specific value that triggers a major maintenance cycle.

\begin{figure}[tbh!]
	\centering
	\begin{tabular}{c}
		\includegraphics[width=10cm]{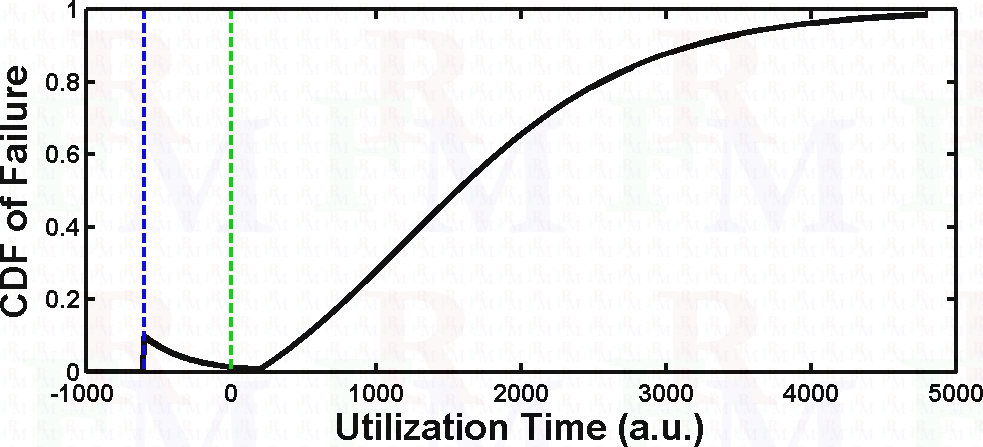}
	\end{tabular}
	\caption{A typical failure profile of a component or system considering the burn-in period. The blue vertical line marks the manufacturing time of the component, while the green vertical line stands for the beginning of its utilization by the user. }
	\label{f-failure_profie1}
\end{figure}

In Figure \ref{f-failure_profie1}, a typical failure profile, which is actually the associated failure CDF of a typical component, is shown. The key behaviors to highlight are burn-in period at the beginning of utilization followed by a {\em sigmoid}-like convergence to most-likely failed state. The burn-in period is illustrated as the time interval between the blue and green vertical lines. Although detailed analysis of burn-in tests is not the focus of this paper, they could have major impacts and consequences on the nature (environment) and society when their Life-Cycle Assessments (LCA) is considered \cite{Tsai2011,Pehnt2008}. 
\textit{}

First, as can be seen from the figure, by adjusting the end of the burn-in period, it is possible to deliver a component at its lowest likely-to-fail state. Although this is of great interest to high performance computing (HPC) applications and is actually practiced, it implicitly implies that a large number of manufactured, working components would not be delivered and are aborted within the burn-in phase \cite{Tsai2011,Tsai2012,Cho2010,Wu2007a}. Despite the fact that this practice increases the reliability of the target HPC applications, it avoids the aborted components from entering into the market while otherwise they could well serve in other applications with low-sensitivity to reliability. This aspect will get more attention in future because it is being argued that the sustainability audit of a system should not only include its immediate action but also the consequences of its indirect actions, such as its influence in the manufacturing and recycling (end of the life) of even those components that have not been used within that system boundary. This approach to the life cycle assessment (LCA) is getting more and more attention in recent years \cite{Pehnt2008,Earles2011}, and it requires proper consideration in system design and operation \cite{Farrahi2014}. In particular, non-invasive approaches to separate high-performance and low-performance manufactured components could be an essential replacement of the brute-force burn-in tests, which also would being a new stream of revenue and creditably for the 'responsible' manufacturers.

\begin{figure}[t!]
	\setlength{\tabcolsep}{1pt}
	\centering
	\begin{tabular}{@{}cc@{}}
				\includegraphics[width=6.5cm]{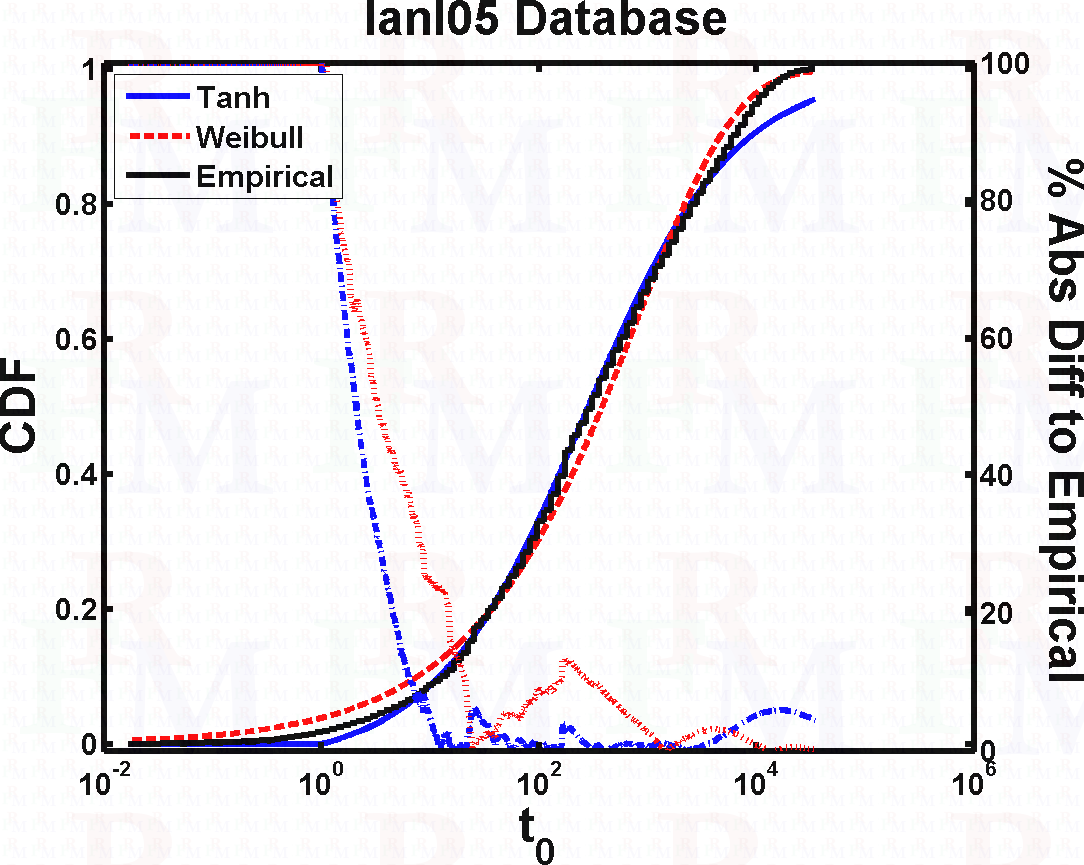} &
		\includegraphics[width=6.5cm]{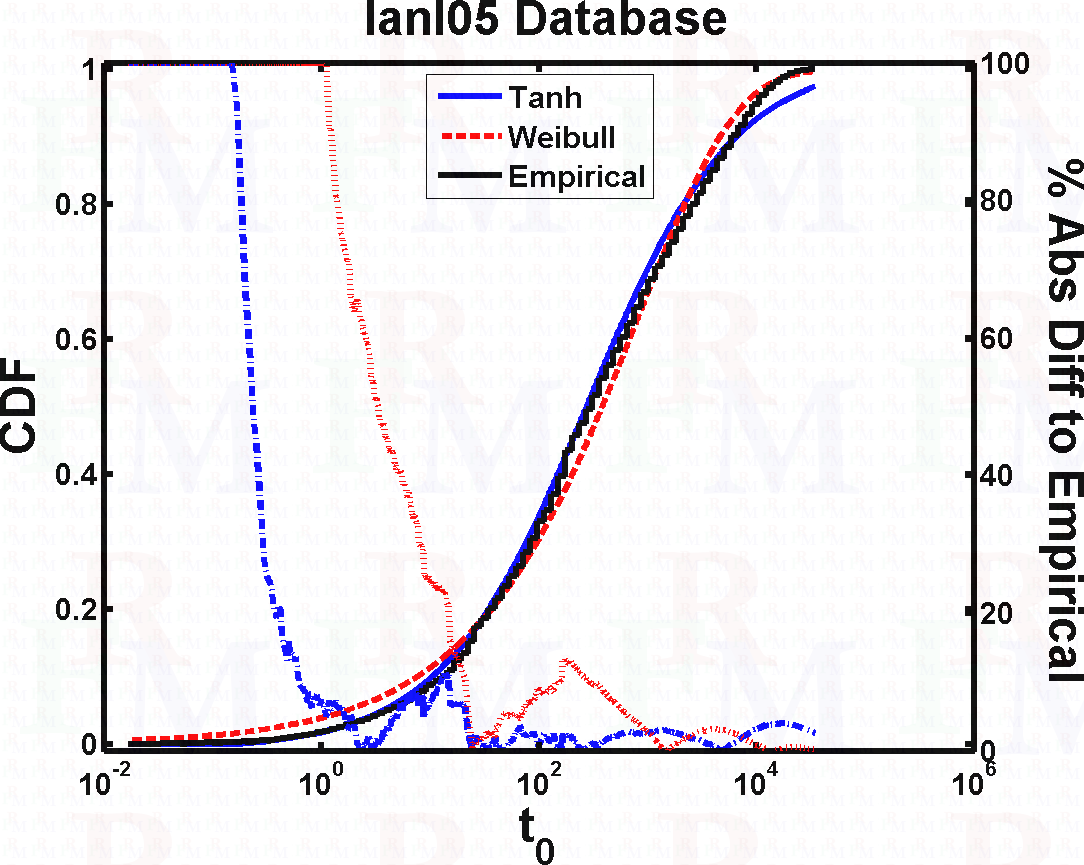} \\
		(a)  & (b) 
	\end{tabular}
	\caption{
		a) The empirical CDF of the {\em lanl05} database compared with its best Weibull and Tanh-Log fits.
		b) The same as (a) but with the c-Tanh-Log distribution. The relative difference percentage between the empirical distribution and the best fits are also shown in each figure.}
	\label{f-CDF-PDF-tanh1}
\end{figure}

In failure analysis of systems with aging and history phenomena, proper selection of failure probability distribution is a critical step. A new Tanh-Log cumulative distribution function was introduced in \cite{Farrahi2012f} with better fitting features. It provides flexibility to absorb various behaviors thanks to its Tanh shape and also nonlinear transformation of the time. In order to remove the limitation of having a zero CDF value at $t=0$, we use a modified version, called the c-Tanh-Log distribution:
\begin{eqnarray}
	\text{CDF}^{x_c, x_s}_{\text{c-Tanh-Log}}(x) = c_{0} +  \frac{1 - c_{0}}{\tanh\!\left(\frac{x_{c}}{2\, x_{s}}\right) + 1}\left\{\tanh\!\left(\frac{x_{c}}{2\, x_{s}}\right) + \tanh\!\left(\frac{\log(x) - x_{c}}{2\, x_{s}}\right)\right\}, \\
	DDF^{x_c, x_s, s}_{\text{c-Tanh-Log}}(C^*) = \frac{s\, \left({\tanh\!\left(\mathrm{log}\!\left(\mathrm{e}^{\frac{x_{c}}{2\, x_{s}}}\, \sqrt{\frac{C^* - 1}{c_{0} + c_{0}\, \mathrm{e}^{\frac{x_{c}}{x_{s}}} - C^*\, \mathrm{e}^{\frac{x_{c}}{x_{s}}} - 1}}\right)\right)}^2 - 1\right)\, \left(c_{0} - 1\right)}{2\, x_{s}\, \left(\tanh\!\left(\frac{x_{c}}{2\, x_{s}}\right) + 1\right)},
	\label{e-c-tanh-log-1}
\end{eqnarray}
where $C^* = \text{CDF}^{x_c, x_s}(x^*)$ at a typical time interval $x^*$, and DDF stands for the Differential Distribution Function, defined as follows, and provides a mean to compare even those CDFs defined on different time intervals \cite{Farrahi2012f}:
$$
\textrm{DDF}^s\Big( C^*=\textrm{CDF}(x^*) \Big) := \frac{1}{s} \frac{\partial \textrm{CDF}(x)}{\partial \log(x)}\bigg|_{x^*}
$$ 
where $s$ is a scaling factor. The term $c_0$ is a constant to absorb the non-zero behavior at the beginning of the utilization phase. To show the performance of the new distribution, the fit of the two distributions on the empirical {\em lanl05} database \cite{Schroeder2010} is shown in Figure \ref{f-CDF-PDF-tanh1}. The time interval of the {\em lanl05} database is considered to be generic, and therefore it is considered with an arbitrary unit (a.u.) of time. The biggest interval in this database is $34,480$ a.u. of time, which determines the interval on all the best fit figures. The empirical CDF of the ({\em union}-interpreted) {\em lanl05} database was retrieved from the Failure Trace Archive (FTA) \cite{Kondo2010}. As can be seen from the figure, the c-Tanh-Log distribution provides a better fit at the beginning. The fitted parameters of the two distributions are  $x_c=5.564 (\pm 0.0035)$ and $x_s=1.577 (\pm 0.0030)$, and  $x_c=5.643 (\pm 0.0019)$, $x_s=1.536 (\pm 0.0012)$, and $c_0=0.01755 (\pm 0.0004)$ respectively. For each parameter value, the $\%95$ interval is shown using $(\pm \cdot)$ notation. The best fits of both distributions and also that of the well-known Weibull distribution \cite{Kondo2010} are shown in Figures \ref{f-CDF-PDF-tanh1}(a) and \ref{f-CDF-PDF-tanh1}(b). The empirical CDF is shown in black, while the Tanh-Log and the Weibull distributions are shown in solid blue and dashed red lines respectively. The p-values of fit (compared to the traditional significance level of 0.05) with respect to the Kolmogorov-Smirnov \cite{Kolmogorov1933,Massey1951} and also the Anderson-Darling \cite{Stephens1978} goodness of fit (GOF) tests are 0.4765 and 0.5659, and 0.4969 and 0.5898 for the two distributions respectively. In this work, we use this empirical distribution as our model for the servers. For the switches and network links, no data was available from the database. Therefore, we assume the same empirical data but with higher performance: Availability intervals of the same empirical data multiplied by arbitrary factors of 2 and 5, respectively, were considered and then used to calculate the associated distributions. It is worth noting that the proposed failure analysis, described in section \ref{sec_MC_Analysis}, is completely independent from the distributions considered for the components, and therefore the results can be recalculated when the empirical data or distribution for any specific component become available.

\begin{figure}[tbh!]
	\centering
	\begin{tabular}{c}
		\includegraphics[width=10cm]{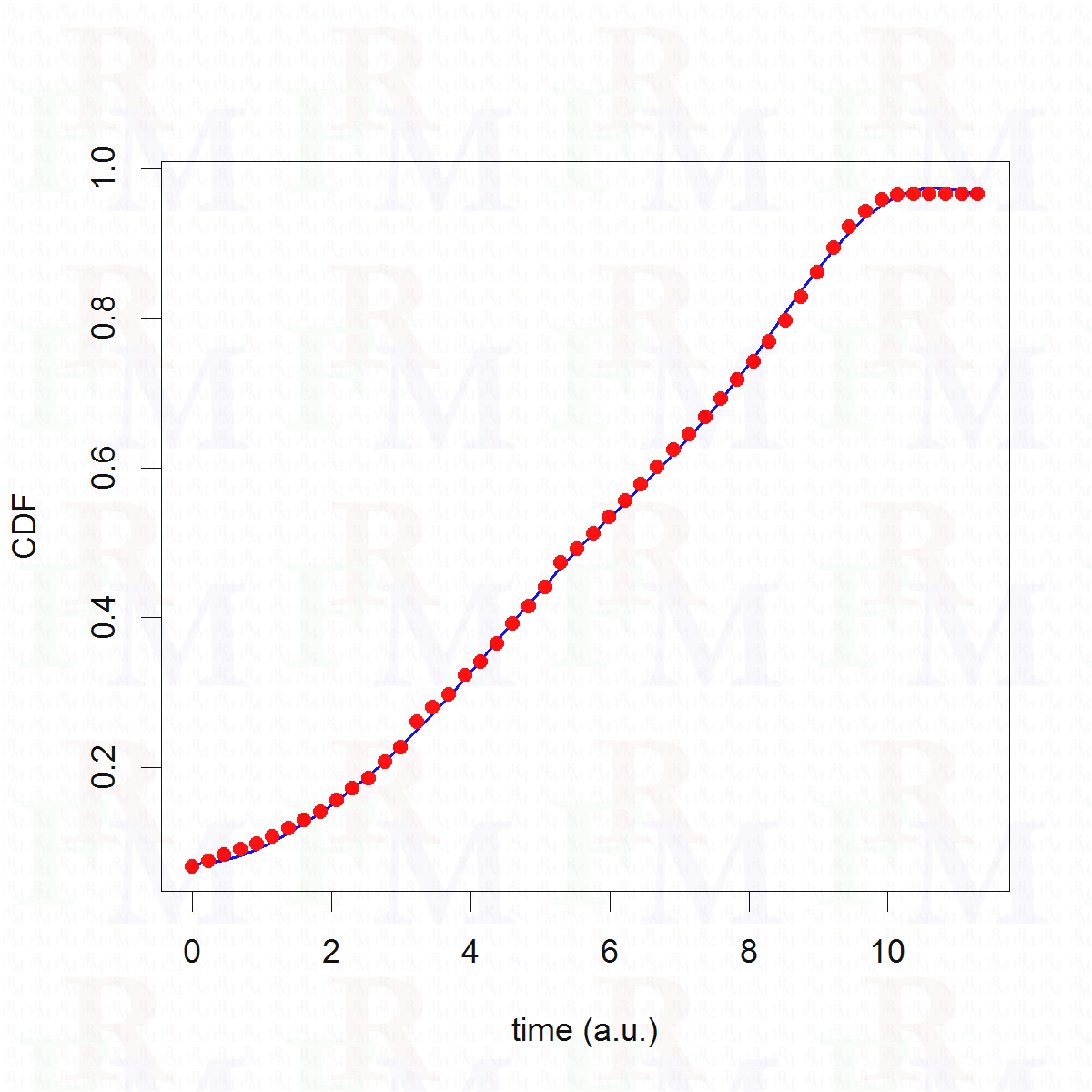}
	\end{tabular}
	\caption{An optimized SVR model fitted on the empirical CDF values associated with the {\em lanl05} database. Red dots are the empirical data and the SVR fit is shown as a continuous curve.}
	\label{f-failure_profie_SVR1}
\end{figure}

\hl{It is worth mentioning that direct statistical methods can be used to model the empirical CDF data without assuming any predefined empirical fitting function. As an example, we use the Support Vector Machine (SVM) and in particular the Support Vector Regression (SVR) }\cite{Nie2014,Cheriet2013}\hl{. In SVM machines, a regression fit is built while the number of samples used from the training data is limited in order to avoid overfitting. The SVR machines are a generalization to the SVM machines in order to extend their scope from classification to regression. The SVR method has been applied to the empirical CDF data obtained from }\cite{Kondo2010}\hl{, and then an optimization on the SVR parameters has been performed. The optimized SVR model achieved a high degree of fitting with as low as an $L_0$ error of 0.011. The optimized model is shown in Figure }\ref{f-failure_profie_SVR1}\hl{. However, considering that the main focus of this paper is on the proposed snapshot-based Monte Carlo method, and also considering that there is only one empirical database currently available, we will perform the validation of the optimized SVR fit in another work in the future with a focus on the statistical modeling of the failure phenomena.}

\section{Modular Data Centers (MDCs)}
\label{sec_MDCs}
In this paper, we consider three typical DC network topologies found frequently in the literature, namely, FatTree \cite{Fuentes2012,Navaridas2010}, BCube \cite{Guo2009}, and MDCube \cite{Wu2009}.\footnote{	MDCube stands for Modularized Data center Cube \cite{Wu2009}.
}
It is important to note that herein we do not claim that these topologies are the best, instead, we have chosen these topologies as they are popular network topologies in modular DCs. The detailed characteristics and design of the mentioned topologies is provided in the following subsections.

\begin{figure}[tbh!]
	\begin{tabular}{cc}
		\includegraphics[width=7cm]{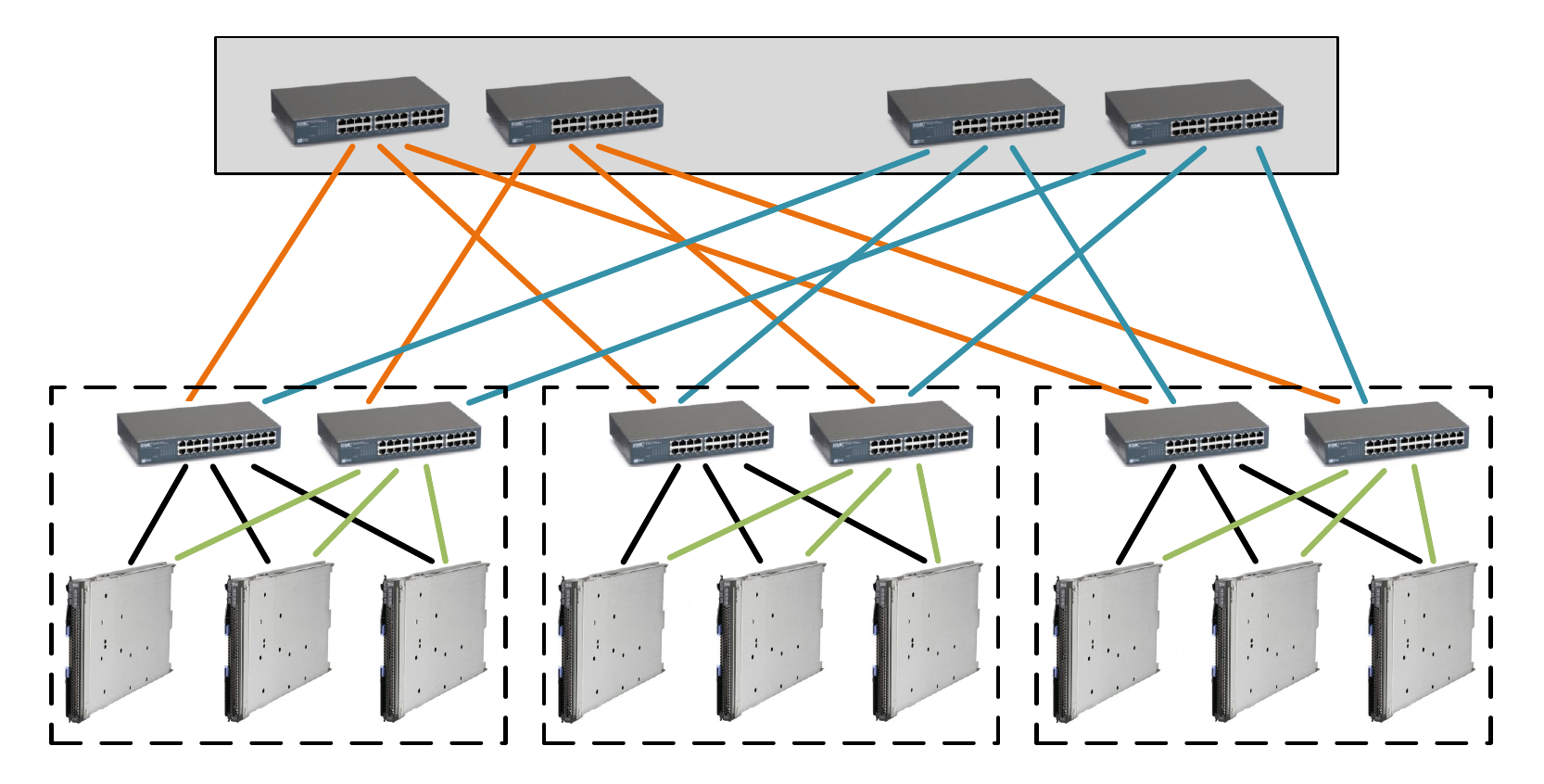} &
		\includegraphics[width=7cm]{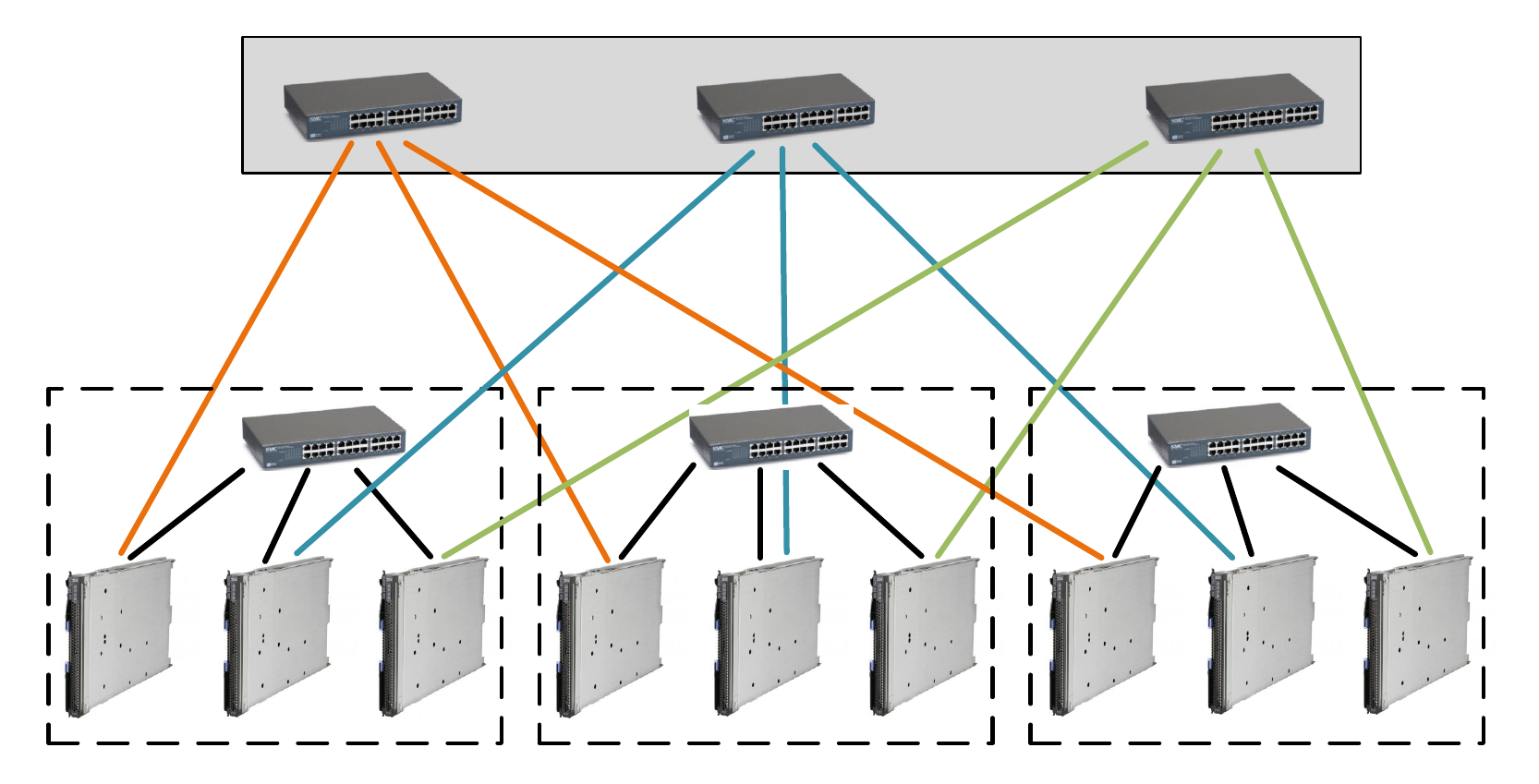} \\
		(a) & (b) 
	\end{tabular}
	\caption{Two illustrative samples of a) FatTree and b) BCube topologies.}
	\label{f-fattree-schematic1}
\end{figure}

\subsection{Fat-Tree Topology}
\label{ssec_fattree_topo}
The FatTree network topology is a special example of the Clos network \cite{Clos1953}, and it has been widely used as a topology for DC networks \cite{Fuentes2012,AlFares2008,Petrini1997}. 
Herein, we refer to FatTree as the cut-down version of the classic FatTree topology proposed by Navaridas et al. in \cite{Navaridas2010}. 
In this topology, the authors utilized the concept of FatTree, to define a DC network topology organized by reducing the number of upward ports of all switches.\footnote{It is worth noting that we do not separately consider the VL2 topology proposed in \cite{Greenberg2009} because of its similarity to the  FatTree topology.} As shown in Figure \ref{f-fattree-schematic1}(a), the considered FatTree topology has two main levels, i.e., the core and the pods. The core level is composed of switches that interconnect the pods in such a way that each switch in the core is connected to at least to one switch in each pod. Each pod is composed of servers and aggregation switches. In fact, core switches are connected to the servers in each pod through the aggregation switches. Finally, each aggregation switch is connected to a different set of servers in a pod. 

We follow the  $K$:$P$-ary $3$-tree FatTree topology with $L$ pods \cite{Navaridas2010}, which means we assume that there are $L$ number of pods (nPod), and each pod contains $K$ number of servers per pod (nSrvPod) with $P$ ports on each (pSrv). Consequently, the network has $L\times K$ number of servers (nSrv), and each pod has $P$ number of aggregation switches each with $K+P$ number of ports (pSwc). In core level, there are $2\times P$ number of core switches each with $L$ number of ports. Figure \ref{f-fattree-schematic1}(a) shows an example of FatTree topology with $L=\text{nPod}=3$, $K=\text{nSrvPod}=3$ and $P=\text{pSrv}=2$.

\subsection{BCube Topology} 
\label{ssec_MDC-BCube}
The BCube topology was proposed as a high-performance and robust network architecture to be used to build MDCs \cite{Guo2009}. 
BCube has a server-centric topology, and it provides high network capacity for various typical DC applications such as MapReduce \cite{Dean2004}. 
Basically, the BCube topology is recursive and has a level-based structure. BCube topology is composed of two types of components, i.e., servers with multiple number of ports, and switches connecting a constant number of servers.     
In the BCube structure, multiple layers of cheap and mini Commercial-off-the-Shelf (CotS) switches could be used to connect the servers. 
In this topology, multiple parallel short paths are provided between any pair of servers. This not only provides high network capacity for all types of traffic, but also greatly improves redundancy, fault tolerance, and load balancing. Finally, in the BCube topology, all networking components in each level are identical, and therefore it would avoid expensive switches with high port density in higher topology levels.

A BCube topology is denoted as BCube (level-$i$), where $i$ is its number of levels. A BCube (level-0) is composed of simply a single switch with $K$ ($=pSwc=nSrvPod$) ports connected to $K$ $P$-port ($=pSrv$) servers. A BCube (level-1) is constructed from $N$ ($=nPod$) number of level-0 cubes and $N$ number of switches with $K$ ports.\footnote{Note that $K$ is also the number of servers in each level-0 cube (nSrvPod).} 
Each switch is connected to all $N$ BCubes (level-0) networks through its connection with one server of each BCube (level-0). Similarly, the general design of a BCube (level-$L$) topology is composed of $N$ number of BCubes (level-$L-1$) and $N$ number of switches with $K$ ports. Again, each switch of BCube (level-$L$) is connected with one server of each BCube (level-0). It is worth to mention that each server in a BCube (level-$L$) has $L + 1$ ports, which are connected to all levels from level-$0$ to level-$L$. Moreover, a BCube (level-$L$) has $N\times K$ total number of servers (nSrv), and $N\times (L+1)$ number of switches (nSwc). Figure \ref{f-fattree-schematic1}(b) shows a BCube (level-1) with $K=3$, $N=3$, $P=2$.

\subsection{MDCube Topology}
Due to the special potentials of the BCube topology, such as high network capacity and directly built from CotS switches), it is a natural step to use the BCube topology as the building block for mega-size DCs \cite{Wu2009}. However, the BCube topology does not have enough agility, and cannot directly scaled to millions of servers just by adding more number of ports to servers and considering more number of CotS switches to the network topology. The major limitation within the design of BCubes is that the number of inter-cube cables required by BCube topology increases linearly with the increasing of the total number of servers.\footnote{It is worth to mention that we can find similar argument for the FatTree network topology.} Accordingly, the MDCube network topology has been proposed as an extension to the BCube topology by utilizing the high-speed interfaces of CotS switches to interconnect multiple containers, each container is a BCube topology by itself.  
From the engineering point of view, the MDCube topology design is interesting from at least two important aspects. 
First, it interconnects BCube containers using high-speed interfaces (10 Gbps) on the BCube's CotS switches. Note that such switches are considered commercial switches today, and having for example 48 number of 1-Gbps ports and a number of 4 10-Gbps ports is the norm. The MDCube topology uses the 10 Gbps ports for its hypercube connections. 
Second, the MDCube topology design can be used to create containers (clusters) of $1000$-$3000$ servers with a high intra-connect capacity, and then interconnect them by a lower bandwidth network.\footnote{By low bandwidth we mean 10 Gbps.}
Note that such a designed feature is practical because various application cases usually require not more than a few thousands servers in the same pool. 

The MDCube topology can be 1-dimensional (MDCube1D) or 2-dimensional (MDCube2D).
The design of MDCube topology is based on treating each BCube as a hypernode and each switch in the BCube as a hyperport. Then, a hyperport from each BCube is connected to a hyperport from another BCube in the same dimension. An example of a 2-dimensional MDCube network topology is shown in Figure \ref{f-mdcube-schematic1} built from $16$ BCube (level-1) containers each one with $K=\text{pSwc}=2$ and $N=\text{nPod}=3$ (see \ref{sec_MDCs} for details). 

\section{The proposed modified MDC topologies: FatTree$^{E}$, BCube$^{E}$, and MDCube2D$^{E}$}
\label{sec_star-topo}
The three topologies considered suffer from limitations in terms of the number of servers and also on the switches in terms of their number of ports. In particular, these topologies require various switches types with different number of ports for their different levels. At the same time, and for example for the BCube topology, the number of pods, i.e., the BCubes at level-0, are restricted to the number of ports of switches. Although these constraints are bearable in small-size data centers, they can pose considerable restrictions when several thousands of servers and switches with high number of ports are targeted. To remove these constraints, we propose extended and modified forms of these topologies in which ``imaginary'' switches are built using single-size actual switches. This not only reduces the complexity of the design in terms of switch port numbers, it is also helpful in making the designs more flexible in terms of the total number of servers. In the following subsections, the details of these modified topologies, which we consider then as starred version of the original topologies, are provided.

\subsection{FatTree$^{E}$ Topology}
\label{ssec_fattree_star_topo}
\begin{figure}[tbh!]
	\centering
	\begin{tabular}{@{}c@{}@{}c}
		\includegraphics[width=8cm]{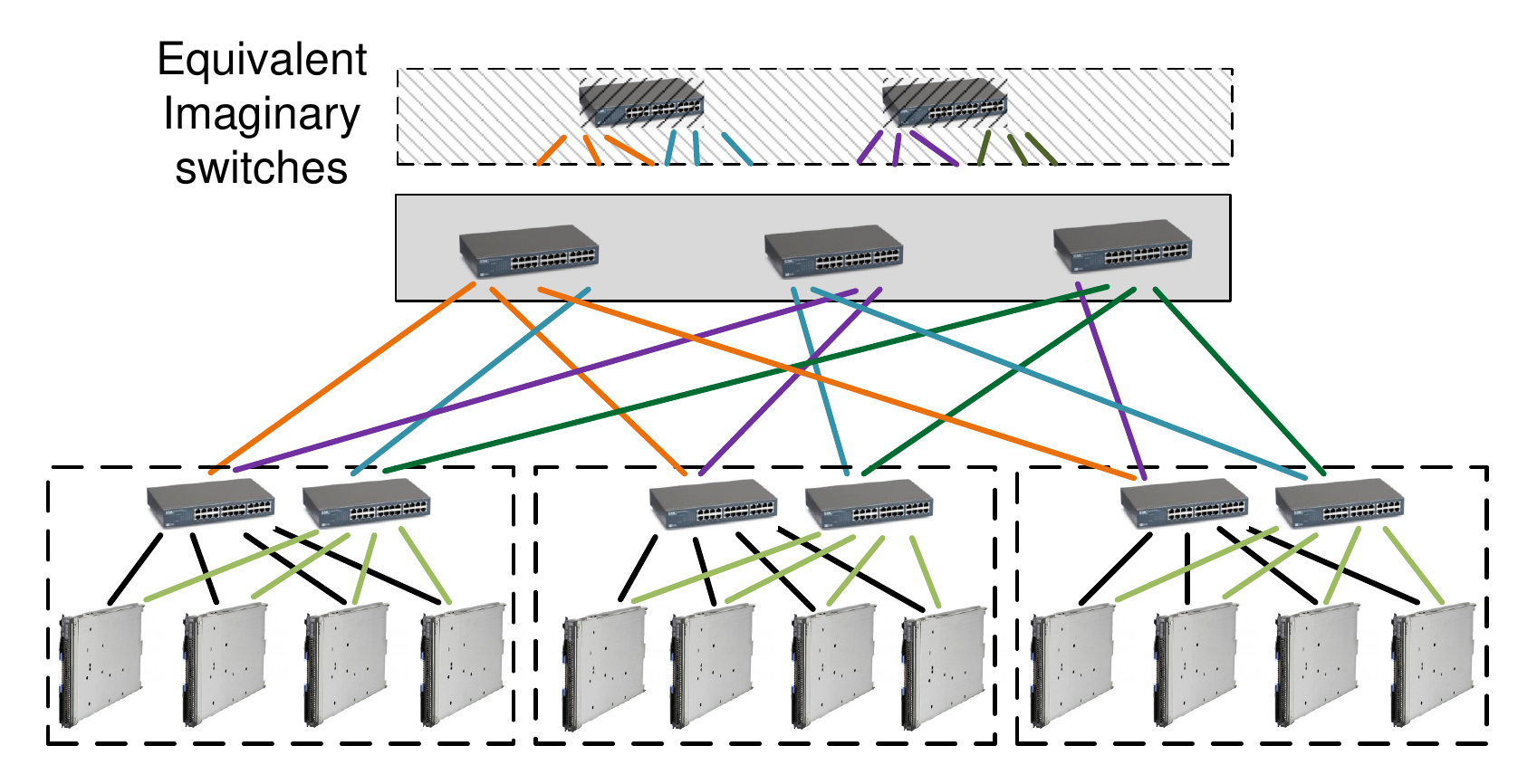} &
		\includegraphics[width=8cm]{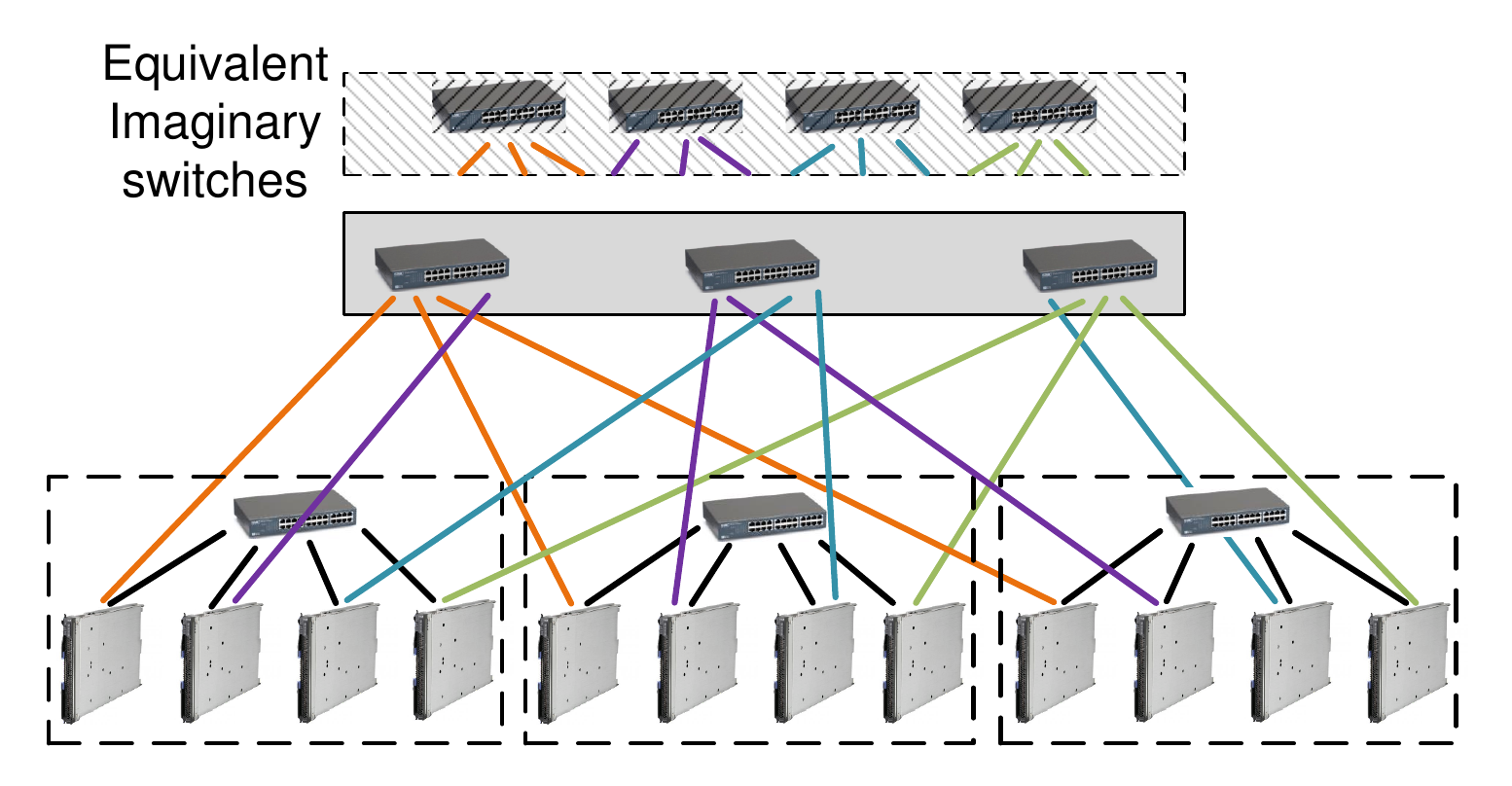} \\
		(a) & (b) 
	\end{tabular}
	\caption{a) A typical FatTree$^{E}$ topology with $L=3$, $K=4$, and $P=2$. The imaginary switches in the level-1 are shown in the dashed area. The connections shown in this area would be those if we considered the FatTree topology. 
		b) A typical BCube$^{E}$ topology with $K=4$, $N=3$, and $P=2$.}
	\label{f-fattreeS-schematic1}
\end{figure}

The main restriction of the FatTree topology defined in section \ref{ssec_fattree_topo} is requiring high-port number switches at the level 1 and higher. This not only can impose additional requirements in the design, it also can increase the probability of large scale failure in the case of the failure of one of main switches. In our proposed FatTree$^{E}$ topology, we use the same switch types across all levels. The switches' port is assigned to the underlying level in a sequential order. That means that one or more of actual switches collectively participate to create the effect of the required high port number imaginary switch compatible with the original FatTree topology. A switch may participate in one or two imaginary switches. The requirement would be to have more number of ports on the actual switches compared to the number of ports of imaginary switches. For example, consider an original FatTree topology with $L$ pods, and $K$ $P$-port servers per pod, which requires $P$ aggregated $K_P$-port switches in each pod, and $2P$ core switches with $L$ ports. The equivalent FatTree$^{E}$ topology would have the same configuration inside pods ($K$ servers and $2$ $K+P$-port aggregated switches). However, at the core level, we would have a $\left[ 2PL/(K+P) \right]$ number of $K+P$-port switches that emulate the functionality of the original high-port $2P$ switches.\footnote{The $\left[ \cdot \right]$ gives the maximum integer less than or equal to its argument.}
The FatTree$^{E}$ topology has the advantage of using the same switching equipment in all levels. An example for $L=\text{nPod}=3$, $K=\text{nSrvPod}=4$, and $P=\text{pSrv}=2$ is shown in Figure \ref{f-fattreeS-schematic1}(a). The imaginary core switches are shown in hashed boxes. It is worth noting that the FatTree$^{E}$ topology is not completely equivalent to its FatTree counterpart. Especially, the fact that imaginary switches are broken on the borders of the actual switches should be mentioned. We will analyze and also propose other modified FatTree$^{E}$ topologies in future to address these aspects, particularly using intra-level connectivity among switches using possibly their extra high-throughput ports.

\subsection{BCube$^{E}$ Topology} 
\label{ssec_BCube_star_topo}
The same strategy, as that used to define the FatTree$^{E}$ topology, is used to generalize BCube to BCube$^{E}$. In the proposed topology, the Level-1 switches are built using the same switch type used in the level-0 cubes. Especially, in the original BCube design, when high-port switches are used in level-0 cubes in order to scale the design, the number of these cubes is required to set to the same number of ports that is a strict restriction on the design. In contrast, in the BCube$^{E}$ design, we propose to choose the number of level-0 cubes independent from the number of ports of switches. Then, the level-1 switches are built using the same switch type of the level-0 cubes with an additional condition that the sum of the number of their ports should be equal to the number of servers in all level-0 cubes. The result of this condition is that we use the same number of switches as that of the level-0 in the level-1 (and higher levels if considered). Compared to the original BCube topology, we consider imaginary switches at the level-1 that would have the same number of ports equal to the number of cubes. 
Similar to the FatTree$^{E}$ topology, the imaginary switches of the BCube$^{E}$ topology are built of parts of consequential switches that are actually used in a level. An example of a BCube$^{E}$ topology with $K=\text{pSwc}=4$, $N=\text{nPod}=3$, and $P=\text{pSrv}=2$ is shown in Figure \ref{f-fattreeS-schematic1}(b), where $K$ is the number of ports of switches (which is also equal to the number of servers in each level-0 cube), $N$ is the number of level-0 cubes, and $P$ is the number of server ports. It is worth mentioning that if original BCube topology was considered, in this example, possible configurations would be 9 (using 3-port switches) and 16 (using 4-port switches) servers. This is because of restriction to have the number of level-0 cubes equal to the port number of a level-1 switch. When, using the BCube$^{E}$ topology, we were able to design a network with 12 servers. This shows the flexibility of the proposed topology in terms of scale and the total number of servers. This effect is more noticeable when switches with high number of ports are considered (for example, in the use case of 3072-server topologies of Section \ref{ssec_usecase_3072}). 

\subsection{MDCube2D$^{E}$ Topology}
\label{ssec_MDCube_star_topo}
The MDCube2D$^{E}$ topology is simply defined by using the BCube$^{E}$ topology as the building block of the original MDCube2D topology. However, the secondary impact of this change in design would be that of high port-number switches that can be easily used to interconnect the BCube$^{E}$ containers (each container is an individual BCube$^{E}$). If $K(=\text{pSwc})<N(=\text{nPod})$ is considered for the MDCube2D$^{E}$' containers, the ratio of the number of containers decreases, and therefore interconnect wiring complexity is improved. In this case, a higher number of links (wiring) among the containers would be expected. To assign the links among containers, we follow the same cyclic procedure as the original MDCube2D topology, i.e., horizontally (vertically) connecting the next available port of a container to the next to previously-connected container. Because of high number of switches per container (when $K<N$), this process should be repeated until all available ports are connected. This would lead to multiple links (multi-path possibility) between each pair of horizontal (vertical) containers. Furthermore, this redundancy in the number of links among the containers enables us to consider different degree of connectivity along horizontal or vertical directions. In other words, if we choose a less number of links per pair along the horizontal direction, for example, we would have a higher number of columns of containers along that direction. Therefore, there is more flexibility by the possibility to have different number of columns and rows in the arrangement of the MDCube2D$^{E}$ topology.
In contrast, we can select $K>N$ that would result in more number of containers because there will be a less number of servers per container. A single link between each pair of containers, along horizontal and vertical directions, could be expected.

\afterpage{\clearpage}

\newpage
\setstretch{1.25}

\parpic{\includegraphics[width=1in,clip,keepaspectratio]{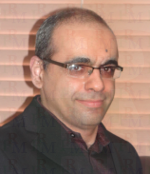}}
\noindent {\bf Reza FARRAHI MOGHADDAM} received his B.Sc. degree in Electrical Engineering and his Ph.D. degree in Physics from the Shahid Bahonar University of Kerman, Iran, in 1995 and 2003, respectively. He has been a Postdoctoral Researcher and a Research Associate with the Synchromedia Laboratory for Multimedia Communication in Telepresence, \'{E}cole de technologie sup\'{e}rieure (University of Quebec) in Montreal (QC), Canada since 2007 and 2012, respectively. He has been a Postdoctoral Researcher with Ericsson Canada since 2015. Reza has published more than 50 technical papers, and has developed several patent-pending inventions. His research interests include sustainability, IoT, drones, behavior analysis, green ICT, green economy, visual perception, and optimization. He is a member of the IEEE.\\

\parpic{\includegraphics[width=1in,clip,keepaspectratio]{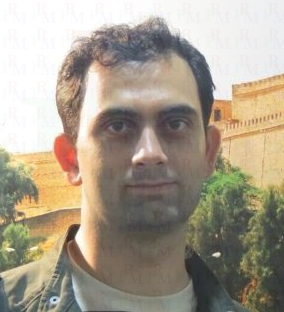}}
\noindent {\bf Vahid ASGHARI} obtained his Ph.D degree in Telecommunications from INRS-EMT at University of Quebec, Montreal (QC), Canada, in 2012 and his B.Sc. degree in electrical engineering from Azad University and the M.Sc. degree in telecommunication system from K.N. Toosi University of Technology in 2002 and 2005, respectively. He is currently working as a postdoctoral research fellow at Synchromedia LAB for Multimedia Communication in Telepresence at \'{E}cole de technologie sup\'{e}rieure (University of Quebec) and at McGill University, Canada. He is a recipient of Postdoctoral Research Fellowship Award from the Quebec Government FQRNT, 2012. He is also co-recipient of Best Thesis Award from INRS University and Best Paper Award from IEEE WCNC 2010.\\

\parpic{\includegraphics[width=1in,clip,keepaspectratio]{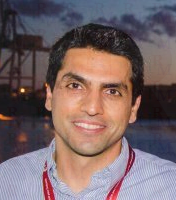}}
\noindent {\bf Fereydoun FARRAHI MOGHADDAM} received his B.Sc. degree in Electronics Engineering from Shahid Bahonar University of Kerman, Iran, in 1999. He obtained his M.Sc. degree in Electronics Engineering from K.N. Toosi University of Technology, Tehran, Iran, in 2001. He joined the Synchromedia Laboratory for Multimedia Communication in Telepresence, \'{E}cole de technologie sup\'{e}rieure (ETS), University of Quebec, Montreal (QC), Canada as a Ph.D. student in September 2008, and received his Ph.D. degree on green distributed networks of data centers in 2014. He is currently working as a Postdoctoral Research Fellow with Synchromedia Lab since 2014. His research interests include IoT, drones, green cloud computing, consistency protocol, smart scheduling, and optimization.\\

\parpic{\includegraphics[width=1in,clip,keepaspectratio]{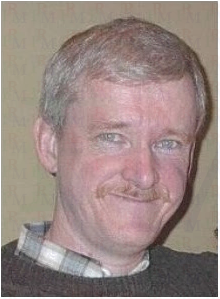}}
\noindent {\bf Yves LEMIEUX} joined Ericsson in 1994 and currently holds the position of Research Engineer in Cloud Technology group in Ericsson Canada Inc. Yves has a number of patents and publications to his credit in the fields of Cellular System Synchronization Selection, Network Resiliency, LTE Core Network Congestion Control, among others. His main interests are now vested in 3GPP based End-to-End QoS and also Virtualization for Cloud Computing. Prior to working at Ericsson, Yves was a Systems Design Engineer at AT\&T Canada and a Radio/Fiber Manager at Rogers Wireless.
Yves received his bachelor degree in Electrical Engineering from the University of Sherbrooke in December 1981, and a Masters Degree in Computer Engineering from Polytechnique in Canada, in June 2005.\\

\parpic{\includegraphics[width=1in,clip,keepaspectratio]{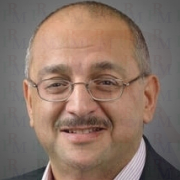}}
\noindent {\bf Mohamed CHERIET} received his B.Eng. from USTHB University (Algiers) in 1984 and his M.Sc. and Ph.D. degrees in Computer Science from the University of Pierre et Marie Curie (Paris VI) in 1985 and 1988 respectively. Since 1992, he has been a professor in the Automation Engineering department at the \'{E}cole de technologie sup\'{e}rieure (University of Quebec), Montreal, and was appointed full professor there in 1998. His interests include document image analysis, OCR, mathematical models for image processing, pattern classification models and learning algorithms, as well as perception in computer vision. Dr. Cheriet has published more than 250 technical papers in the field, and has served as chair or co-chair of the following international conferences: VI'1998, VI'2000, IWFHR'2002, ICFHR'2008, and ISSPA'2012.

\afterpage{\clearpage}
\clearpage

\renewcommand\appendixname{ Supplementary Material }

\appendix

\setcounter{page}{1}
\setcounter{table}{0}
\setcounter{figure}{0}
\setcounter{section}{18}
\renewcommand{\thefigure}{S-\arabic{figure}}
\renewcommand{\thetable}{S-\arabic{table}}
\renewcommand{\thesection}{Supplementary Material \Alph{section}}
\renewcommand{\thesubsection}{Supplementary Material \Alph{section}.\Alph{subsection}}

\section{ }

\subsection{Importance of Modular Data Centers (MDCs) for Deployment of Data Centers in Canada's North}
\label{sec_suppl_mat}
\begin{figure}[tbh!]
\centering
\begin{tabular}{@{}c@{}@{}c@{}}
\includegraphics[height=6.0cm]{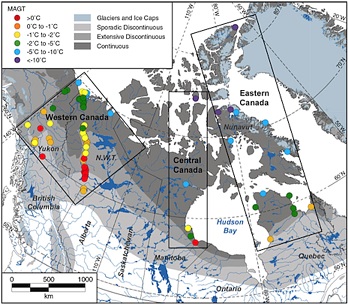} & \includegraphics[height=6.0cm]{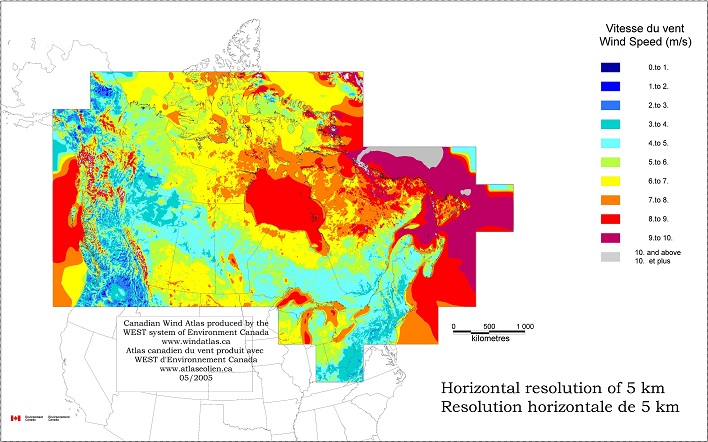} \\
(a) & (b) \\
\includegraphics[height=6.0cm]{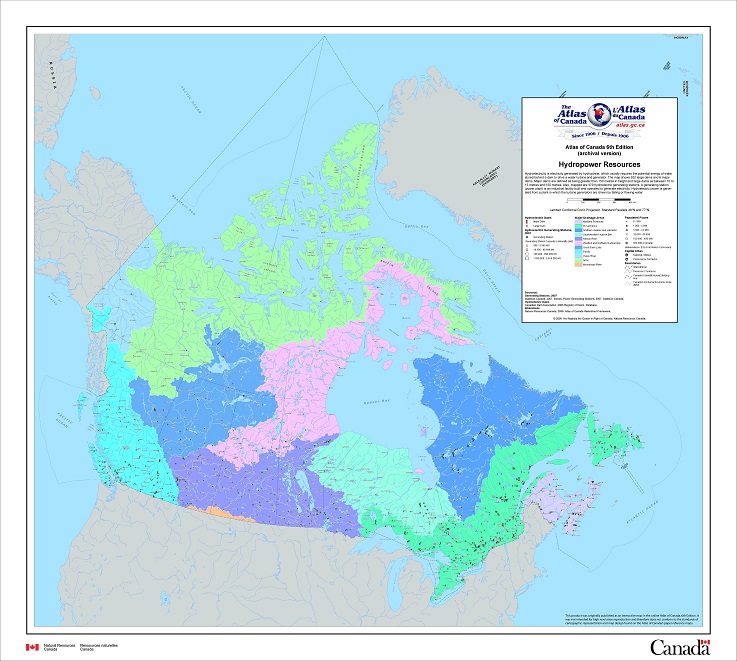} &
\includegraphics[height=6.0cm]{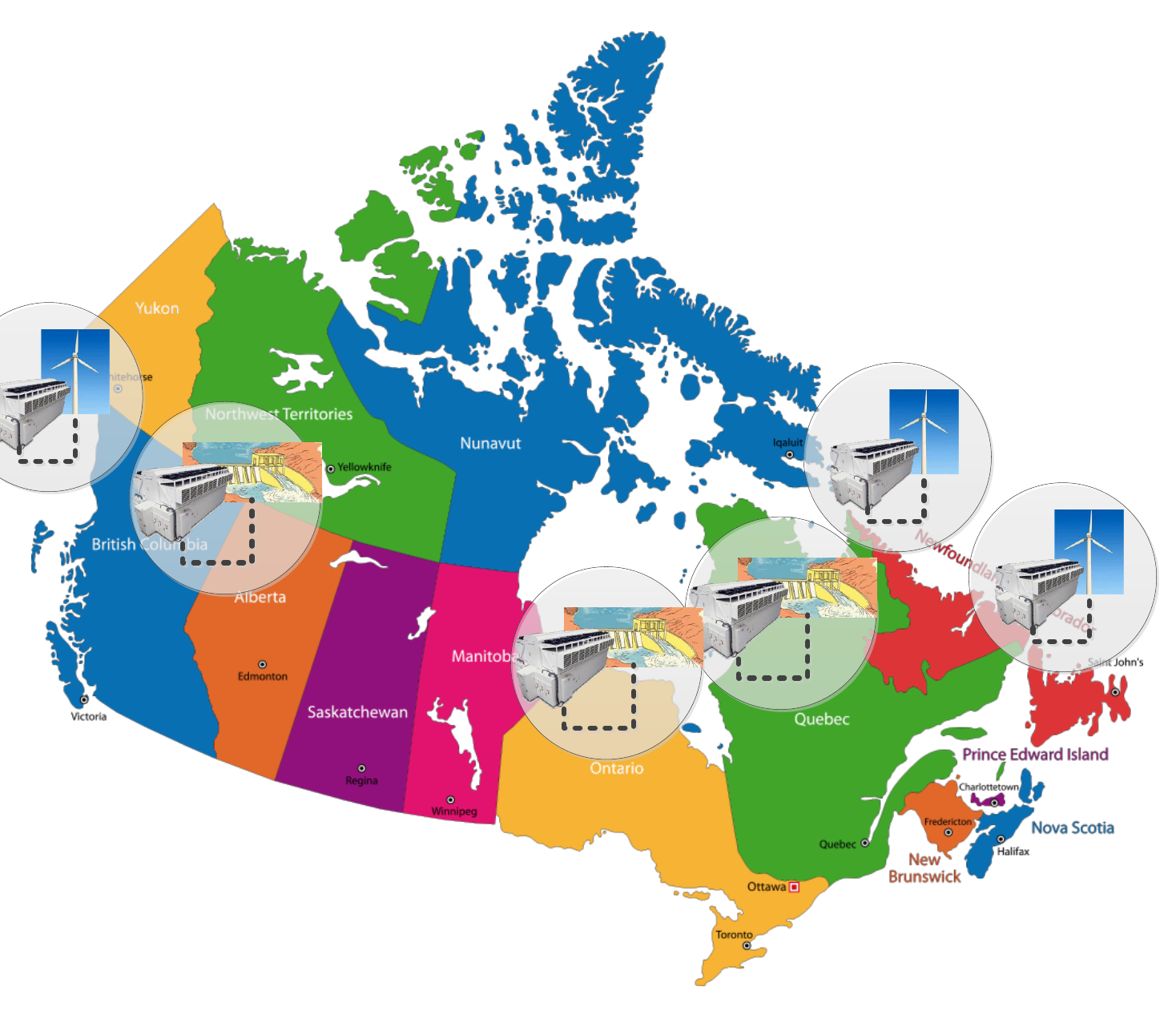} \\ 
(c) & (d)
\end{tabular}
\caption{The potential of Canada's North in terms of low temperature, hydro-electricity, and wind-electricity. a) The mean annual near surface ground temperature in Permafrost of Canada's North (Source: The Canadian Cryospheric Information Network (CCIN)).
b) The annually-averaged wind speed at 50m above ground of the Canadian territory (Source: Canadian Wind Energy Atlas).
c) Canada's major drainage areas. There are 632 large dams and 6 major dams operational with many other in planning/deployment phases (Source: Natural Resources Canada; Earth Sciences Sector; Canada Centre for Mapping and Earth Observation).
d) A proposed schematic configuration of modular, off-grid, highly efficient, and green network of DCs.}
\label{f-canada_hydro_resource1}
\end{figure}

The modular data centers are highly interesting, especially along with strategic north plans in Canada at both provincial and national levels \cite{Berger2010,Fournier2012,Fiser2012}, which target Canada's natural resources, especially untouched hydro-electric and wind-electric sites in the North \citeS{Hofman2009}, among other technologies. The profitability of modular data centers highly depends on survivability of the modules along time, and also on the quality of that survival, especially at the last mile of their terminal stage. The study of such a behavior has been one of this paper's focuses.
Figure \ref{f-canada_hydro_resource1}(a)-(c) shows some considerable potentials in the Canada's North in terms of natural resources and low temperature.\footnote{A link to the temperature map: 
\cooltooltip{Temperature}{A link to Temperature map}{http://aurora.uwaterloo.ca/home/ccw/permafrost/current}{See Temperature map on the website}{http://aurora.uwaterloo.ca/home/ccw/permafrost/current}, \\
A link to the Hydro map:
\cooltooltip{Hydro}{A link to Hydro map}
{http://atlas.nrcan.gc.ca/data/english/maps/energy/hydropower\_resources\_map.pdf}
{See Hydro map on the website}
{http://atlas.nrcan.gc.ca/data/english/maps/energy/hydropower\_resources\_map.pdf}, \\
A link to the wind map: 
\cooltooltip{Wind}{A link to Wind map}{http://www.windatlas.ca/en/maps.php?field=EU&height=50&season=ANU}{See Wind map on the website}{http://www.windatlas.ca/en/maps.php?field=EU\&height=50\&season=ANU}.} The proposed DC sites shown in Figure \ref{f-canada_hydro_resource1}(d) reflect a potential schematic configuration in terms of hydro- or wind-electricity from the perspective of the authors. However, a detailed site analysis and feasibility study should be performed to determine the optimal number of sites and their optimal locations which is beyond the scope of this work. 

Another motivation toward MDCs is their high potential in reducing energy (resource) consumption and also environmental footprint \citeS{Farrahi2012b}. In particular, absence of frequent human access to equipment provides an opportunity in design to effectively contain and handle the heat generated in a DC, especially using non-air (liquid) cooling system designs \citeS{Campbell2012}. It is supported by the observation that the most of electricity consumption in the traditional cooling systems is wasted to extract the heat from the hard-to-reach hot spots \citeS{Barroso2009,Chu2004,Marcinichen2010}. Therefore, MDCs arranged as a distributed network of data centers \citeS{Farrahi2012b,Farrahi2011} and empowered with intelligent sensing (for example, using 
intelligent platform management interface, IPMI) could drastically reduce costs, and in turn can be leveraged toward higher profit or better quality of service (QoS). That said, because of requirement to operate in long term, unreliability at the system level could jeopardize the whole benefit of MDCs. In particular, reliability and scalability of DC network topology is of great importance.

The performance of network switches at their upper-bound is of great importance for these designs that rely on reliability of components at their nominal ratings. If a component is designed to handle a specific performance subject to assumption of a ``probable'' statistics, it can result in the partial failure of system when the system pushes to achieve higher performance that may violate the assumed statistics. For switching components, this can be observed as choking phenomenon when all their ports are used.

Traditionally, switching components have been designed as application-specific integrated circuit (ASICs) that guarantees a designated performance independent from peripheral factors, and also makes them proprietary equipment. However, this vision has been gradually shifted toward more open architectures. One motivation has been the move toward software defined networks (SDNs) whose OpenFlow \citeS{Bianchi2014} is a well-known implementation for layers 2 and 3. Other approaches to SDNs have been Programmable switches via API, virtual network overlays and even other open vswitch approaches. In addition, routing-optimized Linux servers in the form of Open Network Installer Environment (ONIE) have been considered to separate completely hardware and software parts of switching components that in turn provides opportunities to build smart and customized switching software and behaviors to enable maximum utilization of physical connectivity in the data center \citeS{Leake2015}. All these approaches are in answer to high level of under utilization of connectivity in DCs, which is problem especially for deadline-sensitive communications and flows \citeS{Vamanan2012,Zats2012}. 

Although these flow-control mechanism are by definition topology-insensitive, the topology of the DC network could affect the performance especially when multi-path protocols are considered. In this work, we assume that ideal deadline-aware mechanisms are in-place at a higher level, and therefore the associated deadline failure is ignored. The effect of topology on the performance of these mechanisms will be studied in future work. In terms of impact of failure on the multi-path capacity of a design, average max flow among surviving nodes will be studied in this work.

\setcounter{section}{18}
\section{References}

\begingroup
\bibliographystyleS{plain} \bibliographyS{imagep}
\endgroup

\end{document}